\documentclass{article}

\usepackage[english]{babel}

\usepackage[letterpaper,top=2cm,bottom=2cm,left=3cm,right=3cm,marginparwidth=1.75cm]{geometry}

\usepackage{amsmath}
\usepackage{graphicx}
\usepackage[colorlinks=true, allcolors=blue]{hyperref}

\title{Towards the Unmanned Aerial Vehicle Traffic Management Systems (UTMs): Security Risks and Challenges}
\author{Konstantinos Spalas \\
Department of Informatics and Telecommunications \\
M.Sc. in Computer Science\\
University of the Peloponnese (UOP) \\
Tripoli, GR-23131, Greece \\
dit2318cst@uop.gr}
\date{February 2024}
\providecommand{\keywords}[1]{\textbf{\textit{Key Words:}} #1}
\begin{document}

\maketitle
\clearpage
\listoffigures
\listoftables
\tableofcontents
\clearpage

\begin{abstract}   

Every aspect of our life depends on the ability to communicate effectively. Organizations  that  manage to establish communication routines, protocols and means thrive. An Aerial Traffic Management System operates similarly as an organization but certainly in a more strict manner. Third party agencies ensure several aspects of their functionality,  the utmost to be consider \textit{safety}. Many people take safety as granted but it is a pretty difficult part our daily functions. Thus, apart from digesting new things and habits of the new era, simultaneously we have to ensure safety in every part of it. \par
It is true that the more data we  produce, the more information we create and the more specialization we must introduce in order to be effective in a reasonable time basis. A Unmanned Aircraft System Traffic Management (UTM) is a system that consists of miscellaneous modules where each of them needs its consideration regarding safety. In other words, a UTM is the state-of-the-art system that demand a high quality of  services and specialization, if we need to consider them reliable.

\end{abstract}

\section{Intro}
\subsection{History of Aviation}
During the early stages of aviation, the communications that used were visual signal using piece of colored strings or airplanes' maneuvers that were related to certain preset words. There was the assumption that the possibility of more than one airplane could fly on the same terminal was very low. Thus, it was impossible that two planes would collide. Nevertheless, in 1956 two planes  crashed over the Grand Canyon. This incident ignited several actions in order to keep flights in safety, regarding the collaboration and introducing one central authority that manages the airspace.\par
Aviation has passed from several stages in general, in respect of the technological status. At the beginning of the twentieth century , the Wrigth brothers managed to create some gliders, evolving simultaneously the area of aviation studying the aerodynamic science. It is widely known that aviation and aerospace uses the state-of-the-art technology. Apart of the structure, the aerodynamics, the size, great evolution has being taken place over the communication and data exchange sector. In the early years it was highly demanded that pilots made test or normal flights, in order to declare emergencies or communicate with the base, had to transmit critical data for performance.  As the field of aviation took a lot of notice and as its capability of monitoring groundfields without getting noticed, part of aviation got militarized. In WWI, Military Air Forces played critical role to the war's outcome because the countermeasures for that kind of attacks was in the very early stages.
\subsection{Communications and Navigation Guidance in Classical Aviation}
Radio communication was first used in aircraft just prior to World War I (WWI) \cite{1919telephony}. The first airborne radios were in zeppelins. Nevertheless,  military needs triggered  development of light radio sets that could be mounted on the  aircrafts and  they could report their observations immediately. The first experimental radio transmission from an airplane was conducted by the U.S. Navy in August 1910. The first aircraft radios transmitted  using radiotelegraphy, sending Morse code. This configuration required two-seat aircraft, while the backseat  pilot used the telegraph to transmit messages. During WWI, Amplitude Modulation (AM), which a way to transmit data in analog manner, made capable one pilot  to transmit messages while operating the airplane, eliminating  the need of a second crew having this duty. That led to lighter configuration leaving space for machine guns or third party equipment. \par
Radios are based on radio waves which  are electromagnetic pulse  and part of the electronic spectrum. The atmosphere is filled with that kind of waves. Each wave occurs at a specific frequency and has a corresponding wavelength (period). The relationship between frequency and period is inversely proportional. A high-frequency wave has a short wave length and a low frequency \textit{f} wave has a long period \textit{T}.
\begin{equation}
    f=1/T \iff  T=1/f
\end{equation}
 In order to perform communication over two stations, some hardware must be used to manipulate the multi-band microwaves that carries the information transmitted. Such devices are the transmitter,  the antennas and the receiver.\par
A \textit{transmitter} consists of a precise oscillating circuit or oscillator that creates an Alternate Current (AC) carrier wave frequency. This is combined with amplification circuits or amplifiers. The distance a carrier wave travels is directly related to the amplification of the signal sent to the antenna. 
The \textit{antennas} are  simply conductors of lengths proportional to the wavelength of the oscillated frequency put out by the transmitter. An antenna captures the desired carrier wave as well as many other radio waves that are present in the atmosphere. Finally, a \textit{receiver}  isolates the desired carrier wave with its information.\par
Currently,  pilots and the ground control tower are mainly communicate via UHF (Ultra High Frequency, around 1 GHz ) or VHF (Very High Frequency, around 100-300MHz) analog voice radios ~\cite{analogtodigital}. When an analog voice radio communication technology is used, all pilots in the same sector in order to communicate with an air traffic controller must be tuned on the same frequency. This can be challenging considering the expected air traffic growth. Statistical data on air traffic reveals that there is an  increase  trend regarding the  transportation industry over the air. Long term forecast studies provided by Boeing predict a \textit{5\%} growth rate of the world air traffic load between 2011 and 2030. This growth is due to many factors such as the  more competitive low-cost airlines, the  increased passenger demand and the greater need for companies to provide a better service to their customers. Nowadays, the air traffic load is still increasing, leading to a congestion of the worldwide analog voice frequency allocated to the civil aviation.

The term “data link”  is commonly used among the civil and military aviation community as the digital communications between an aircrafts and a ground stations (A2G) or between aircrafts to aircrafts (A2A). There are several benefits for digital data exchange to be preferred. For instance, using digital way to transmit signals we are able to:
\begin{enumerate}
    \item Confirm the ground instructions aircrafts receive via special on board devices.
    \item Correct errors during transmission. Thus, no data loss due to signal jamming.
\end{enumerate}

Another paradigm of digital communication is the utilization of the satellites. Aircrafts communicate with satellites for both operational an non operational services~\cite{sat}. The majority role is for safety reasons using L-Band SATCOM services. L-band Digital Aeronautical Communications System (LDACS)~\cite{ldacs} is one of the radio access technologies revealing the future aeronautical communication infrastructures that will allow aircraft to be digitally connected to the Aeronautical Telecommunication Networks (ATN) during all phases of flight. Specifically, LDACS shall connect an aircraft
operating in the  airspace by deploying a network of ground stations, each one of them
covering a part of the airspace. An aircraft carrying an LDACS radio will then be able to connect
to the Airspace Traffic Management System by communicating with the LDACS Ground Station (GS) covering its current location. This kind of deployment   is similar to cellular mobile communication networks, functioned in areas called cells. A Ground  Station operating in LDACS will be capable of serving up to  512 aircrafts \cite{LDACSMAIN}, considering the data exchange in general (guidance and communications).

\subsection{Internet of Things (IoT)}\label{iot}
The Internet of things (IoT) describes devices with sensors, processing ability, software and other technologies that connect and exchange data with other devices and systems over the Internet or other communications networks. Many IoT devices are  embedded with technology such as sensors and software and can include mechanical and digital machines and consumer objects. Increasingly, organizations in a variety of industries are using IoT to operate more efficiently, deliver enhanced customer service, improve decision-making and increase the value of the business. With IoT, data is transferable over a network without requiring human-to-human or human-to-computer interactions. A thing in the internet of things can be a person with a heart monitor implant, a farm animal with a biochip transponder, an automobile that has built-in sensors to alert the driver when tire pressure is low, or any other natural or man-made object that can be assigned an Internet Protocol address and is able to transfer data over a network.
Several fields of embedded systems are wireless sensor networks, control systems, automation (including home and building automation), independently and collectively enable the Internet of things. In the consumer market, IoT technology is most synonymous with "smart home" products, including devices and appliances (lighting fixtures, thermostats, home security systems, cameras, and other home appliances) that support one or more common ecosystems and can be controlled via devices associated with that ecosystem, such as smartphones and smart speakers. IoT is also used in healthcare systems.\par
The Internet of Things  is infiltrating many businesses. It provides simple means to collect and analyze technical system data to identify and optimize the performance of many things in our private and work lives. This technical revolution is also revealing new challenges and issues with our current IoT technologies. New solutions like Artificial Intelligence, Blockchain or 5G promise to overcome these challenges\cite{iot_and_5g}.\par
\begin{figure}[h]
    \centering
    \includegraphics[width=0.7\textwidth]{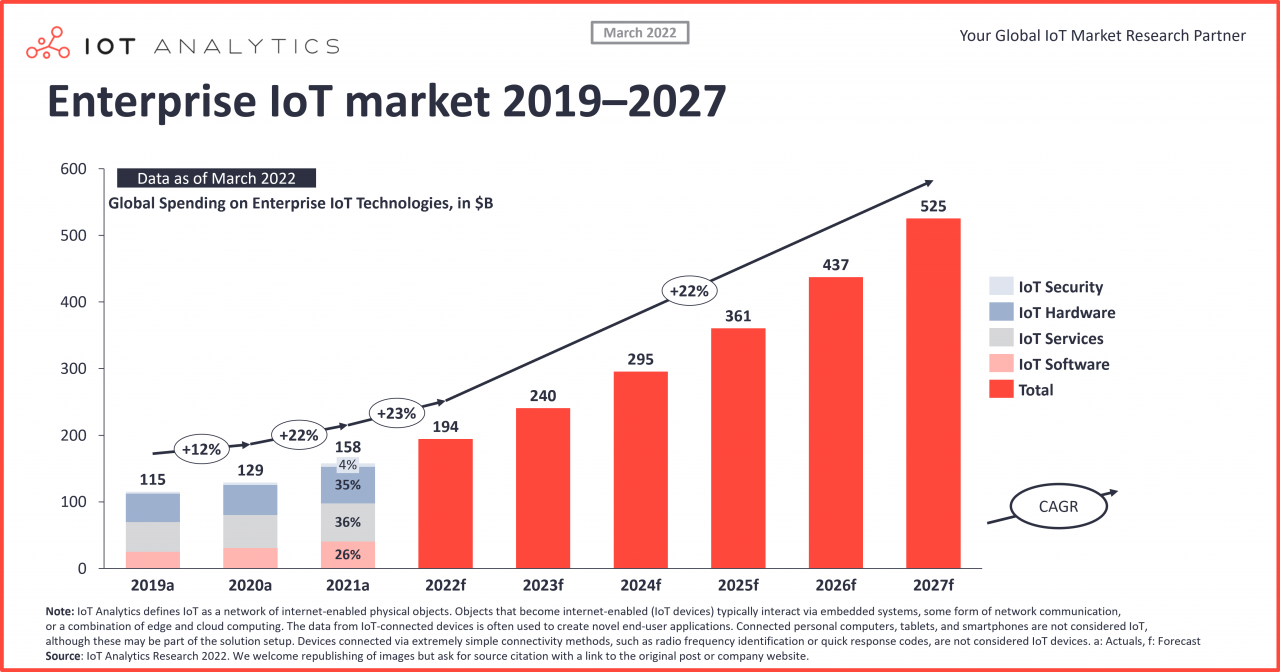}
    \caption{IoT growth}
    \label{fig:iot_grow}
\end{figure}

The enterprise IoT market grew 22.4\% to \$157.9 billion in 2021, according to the March 2022 update of IoT Analytics Global IoT Enterprise Spending Dashboard. The market grew slightly slower than the 24\%, that predicted last year, due to several factors, including a slower-than-anticipated overall economic recovery, a lack of chipsets and disrupted supply chains. North America was the fastest growing region in 2021 (+24.1\%), and process manufacturing was the fastest-growing segment (+25\%).
At this point, IoT Analytics forecasts the IoT market size to grow at a CAGR of 22.0\% to \$525 billion from 2022 until 2027 (fig.\ref{fig:iot_grow}). The five-year forecast has been lowered from the previous year. A number of growth headwinds have had a much more profound impact than previously anticipated, namely supply shortages and disruptions (most notably chip shortages which are now expected to extend well into 2024 and possibly even beyond) and labor shortages, especially for sought-after software jobs. Despite the lowered growth projections, IoT remains a very hot technology topic with many projects focusing to enhance peoples' life. 

\subsection{Unmanned Aerial Vehicles (UAVs) }\label{uav}
An unmanned aerial vehicle (UAV), commonly known as a \textit{drone}, is an aircraft without any human pilot, crew or passengers on board. UAVs were originally developed through the twentieth century for military missions. As control technologies improved and costs fell, their use expanded to many non-military applications. These include aerial photography, precision agriculture, forest catastrophe monitoring, environmental monitoring, policing and surveillance, infrastructure inspections, smuggling, product deliveries, entertainment and many others.\par
Various terminologies are employed to refer to unmanned aircraft. Some of them include: UAV
(Unmanned Aerial Vehicle), UAS (Unmanned Aerial System), drone, RPA (Remotely Piloted Aircraft), RPV (Remotely Piloted Vehicle), and RPAS (Remotely Piloted Air System). There are a 
 slight variations in the terminologies utilized by different countries and
institutions when referring to unmanned aircraft\cite{uav_names}.\par
The Unmanned Aerial Vehicles Market size is estimated at USD 17.31 billion in 2024, and is expected to reach USD 32.95 billion by 2029, growing at a CAGR of 13.74\% during the forecast period (2024-2029)\cite{uav_growth}. While UAV technologies growing, they  have allowed manufacturers to produce a wide range of models in different sizes, weights, and shapes that can carry different sensor payloads, making them favorable across a broad application base. However, the lack of regulations and restrictions on the flying of UAVs beyond the visual line of sight (BVLOS) in several countries across the world has restrained the market's growth to its full potential. Other factors like security and safety concerns and   trained pilots are also anticipated to challenge the growth of the UAV market to a certain extent. (Source: \textit{https://www.mordorintelligence.com/industry-reports/uav-market})\par
UAVs are generally classified by their flying principle because of their aerodynamic structure. These having heavier mass will depend on propulsive thrust to fly into the air and categorized into two types, the \textit{rotor} and \textit{wing} type. Rotor UAVs will depend on multiple rotors and propellers attached to them in order to generate the required amount of thrust to lift upwards. The differential thrust make the capable to make turns, slips and general to manage their orientation. Similarly, UAVs with wing types will depend on their wings to produce the aerodynamic effect for lifting upwards into the air. This is classified further into three sub-categories,  flapping-wing, fixed-wing and flying-wing. The UAV with a light weight like a parachute, balloons and blimps will rely on forces to fly in the air.\par
Floreano et al.\cite{floreano2015science} discussed the different categories of UAV/Drones. Drones are categorized based on their mass and flight time. UAVs with heavier mass will have the capability to carry heavy payload and can perform autonomous and multiple tasks. Fixed-wing and Rotor-type UAVs are heavier in mass and relatively big in structure. Considering the aerodynamic efficiency, fixed-wing UAVs can have more flight time compared to rotor-type.\par
\begin{figure}[h]
    \centering
    \includegraphics[width=0.75\linewidth]{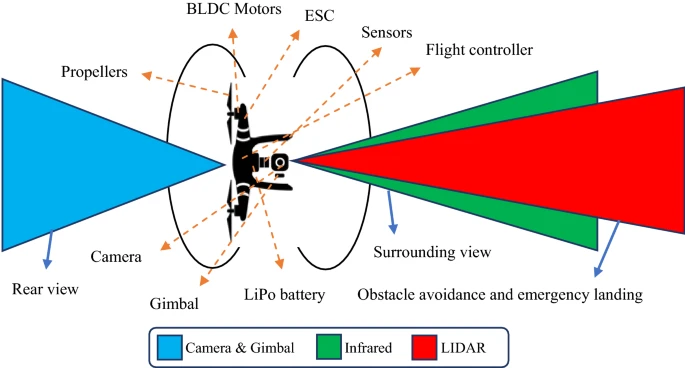}
    \caption{Drone break down}
    \label{fig:drone_break_down}
\end{figure}
\begin{figure}[h]
    \centering
    \includegraphics[width=0.75\linewidth]{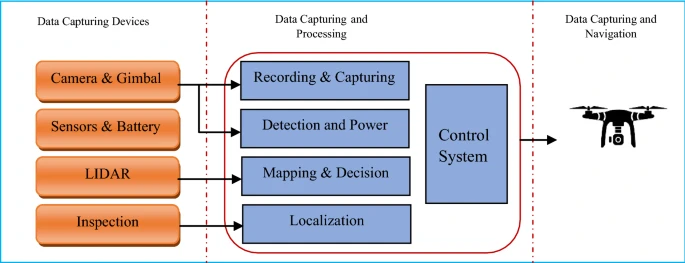}
    \caption{Drone functions}
    \label{fig:uav_functions}
\end{figure}

The hardware in fig.\ref{fig:drone_break_down} reveals the onboard components  used for different applications such as path planning, collision avoidance and inspection during the hovering of UAV. Light detection and ranging (LIDAR) and infrared devices are mainly used for collision avoidance and mapping, whereas camera and GPS are used for surveillance of particular area or path of the UAV in front and rear direction\cite{Ahmed2022-kq}. In fig.\ref{fig:uav_functions} there is a block diagram that distinguish each function of a UAV and their correlation.

\clearpage
\subsection{Authorities Managing Aerial Vehicles Traffic}\label{ch:Regulators}
The Federal Aviation Administration (FAA), which is responsible for the regulation and oversight of civil aviation within the United States, forecasts that the recreational small drone market will saturate at around 1.81 million units over the year 2026 and the commercial drone fleet will likely be at around 858,000 in the US. In a similar context, the number of consumer leisure drones will be about 7 million in 2050 and 400,000 for governmental and commercial purposes in Europe\cite{utm}.
\subsubsection{FAA and UAVs}\label{ch:FAA}
 The Federal Aviation Administration (FAA) is a governing body under the United States Department of Transportation that is responsible for a wide range of regulatory activities related to the United States airspace. In a recently published final rule, the FAA addresses several concerns such as the need for a system to identify all aircrafts flying in national airspace, as well as the implementation of a separate system from the prevalent Automatic Dependent Surveillance–Broadcast system to prevent interference with manned aircrafts. Indicative, FAA is responsible for:
 \begin{itemize}    
 
     \item Directing air traffic in controlled airspace
      \item Ensuring people from space launching
     \item Airport safety/inspection
     \item Standardising airport design, construction
         \item Regulating for flights, inspection standards,
         
 \end{itemize}
 and many others, while is been awake for new challenges and evolution due to technological improvements, implementations, etc. Thus,  FAA  enforces additional policies and certifications that would allow commercial and recreational flights in the US airspace. These are described in detail in \textit{Part 107} published by the FAA that addresses small UAS (sUAS) that weigh less than 55 lbs (24.9 kg).\par
 All unmanned aircrafts  identified by the same document by class and capacity should have a remote identification (RID). RID  broadcasts vital operational information including but not limited to drone's ID, its latitude and longitude information, current altitude, velocity, the ground control station, and the overall status of the UAS along with a timestamp. This information must be broadcast in ways that current wireless systems may recognize, record, and process. \par
 While US governing agencies retain the use of the word UAS for now, the International Civil Aviation Organization (ICAO) terminology is remotely piloted aircraft systems. The FAA describes the RID implementation as a \textit{digital license plate} for all UAS flying in the United States airspace. They outline additional policies including several options for compliance, operating rules, and design and production guidelines for manufacturers. As the September 2023 deadline for compliance draws near, this article highlights possible deployment applications and challenges\cite{faa}. Hence,all pilots will be required to either have RID systems that comply with the described specifications or fly exclusively in designated areas known as FAA-recognized identification areas (FRIAs). 
\subsubsection{EASA and UAVs}
European Union Aviation Safety Agency (EASA) has been the authority that secures the safety of aviation and frames the  protection form in Europe. As an independent and neutral entity, EASA ensures confidence in safe air operations in Europe and worldwide by proposing and formulating rules, standards, and guidance-by certifying aircraft, parts, and equipment and by approving and overseeing organisations in all aviation domains. Thus, EASA aims to:
\begin{itemize}
    \item Ensure the highest common level of safety protection for EU citizens
    \item Ensure the highest common level of environmental protection
    \item Single regulatory and certification process among Member States
    \item Facilitate the internal aviation single market and create a level playing field
    \item  Work with other international aviation organisations and regulators
\end{itemize}
In order  achieve its mission, EASA breaks down the ultimate purpose into several tasks:
\begin{itemize}
    \item     Draft implementing rules in all fields pertinent to the EASA mission
   \item  Certify and approve products and organisations, in fields where EASA has exclusive competence (e.g. airworthiness)
    \item Provide oversight and support to Member States in fields where EASA has shared competence (e.g. Air Operations , Air Traffic Management)
    \item Promote the use of European and worldwide standards
    \item Cooperate with international actors in order to achieve the highest safety level for EU citizens globally (e.g. EU safety list, Third Country Operators authorisations)

\end{itemize}

European Union Regulations  2019/947 and 2019/945 set a framework for drones in order to fly with safety in European airspace. Regulation 2019/947 is applied from 2020 for all the European members (including Norway and Liechtenstein) and is about to expand to  Switzerland and Iceland. It defines three categories of civil drones operations: \textit{open},  \textit{specific} and \textit{certified}.\par
The \textit{open} allows people to operate drones in private  areas that are not interfere formal flights. This category addresses the lower risk for civil drones operations and no special authorization is mandatory before flight.
It is divided in three subcategories (A1,A2,A3).\par
The \textit{specific} covers operation with more risk, whereas the drone operator receives an authorization before the mission.\par
The \textit{certified} refers to flights with high safety risk and thus certification for then drone and its operator is mandatory.

\subsubsection{International Civil Aviation Organization (ICAO)}
The first ICAO exploratory meeting on UAVs was held in Montreal on 23 and 24 May 2006. Its objective was to determine the potential role of ICAO in UAV regulatory development work. The
meeting agreed that although there would eventually be a wide range of technical and performance specifications and standards, only a portion of those would need to become ICAO Standards and Recommended Practices (SARPs). It was also determined that ICAO was not the most suitable body to lead the effort to develop such specifications. However, it was agreed that there was a need for
harmonization of terms, strategies and principles with respect to the regulatory framework.

\subsection{Cryptography}

The goods of Conﬁdentiality, Integrity and data Availability are fundamental
rights for every human. Securing these fundamentals is one of my major aspects.
As we live in a modern, digital, high quality and state-of-the-art environment, information flow in various directions via various paths when we use digital  devices and communication software. Thus, we rely the  security of our data on the  cryptographic systems encapsulated on the above software or hardware.\par
Cryptography, as a method, has its roots from the very ancient years when sensitive information had to be hidden during their transmission. With such practises,people managed to conserve the aforementioned fundamentals. The ﬁrst attempt of a machine that ciphers a message is aged in ancient Sparta . That machine was called Spartan baton. Later, Caesar came up to a mathematical method that relied in letter sifting. He was able to send messages to his generals without these been
broken. \par
While an action creates a reaction, cryptography has its own rivalry
which is cryptanalysis, and vice versa. As long the latter become more eﬃcient and
sophisticated the need for more eﬀective methods of cryptography is a must. While
humanity constantly evolves the ﬁeld of cryptology, which is both cryptography and
cryptanalysis, it has been converted to a remarkable science. Very advanced math
techniques are the basis for both legs of this science. Apart the power of mathematics,
computer hardware plays key role in modern cryptology. A state-of-the-art hardware
are the quantum computers which states the future of cryptography uncertain due
to their high computational power. Thus, eﬀorts to break ciphers will not be that
ineﬃcient anymore, in respect of time and space.

\subsubsection{Symmetric Cryptography}
\begin{figure}[h]
    \centering
    \includegraphics[width=0.75\linewidth]{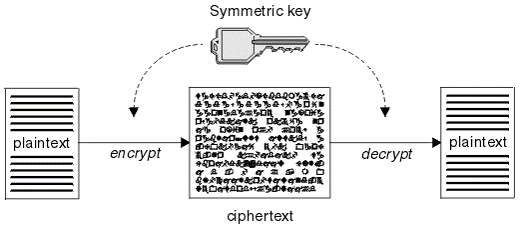}
    \caption{Symmetric key cryptography}
    \label{fig:sym_key}
\end{figure}

Symmetric key cryptography refers to encryption methods in which both the
sender and receiver share the same key(fig.\ref{fig:sym_key}). This was the only kind of
encryption publicly known until June 1976.

Symmetric key ciphers are implemented as either block ciphers or stream ciphers.
A block cipher enciphers input in blocks of plaintext as opposed to individual characters, the input form used by a stream cipher.

In one example of a $One-Time-Padding$ scheme, Alice and Bob both agree on a single, random
n-bit binary vector $p$ (known as the pad). In this case, $p$ is the private key shared by Bob and
Alice. When Alice would like to transmit a message to Bob, she performs a $modulo-2$ addition between $p$ and her message $m$ and transmits the result $r$ to Bob.
\begin{equation*}
    r=m\oplus p
\end{equation*}
Note that binary $modulo-2$ addition is the same as XOR (exclusive OR) . 

This addition constitutes the entirety of the encryption process. When Bob receives Alice’s encrypted message, he
uses the same pad $p$ and performs the same  addition of $p$ to the received message. What it comes is 
\begin{equation*}
    r\oplus p=(m\oplus p)\oplus p=m\oplus (p\oplus p)=m \oplus 0=m
\end{equation*}
, which is 
the original message. Since only Alice and Bob know the secret pad, any third party that
intercepts the encrypted message will have a difficult time deducing the original message. \par
One of the primary disadvantages to private key cryptography relates to the difficulty of
keeping the private key secret or use more than one an synchronize them. In order to protect the cryptosystem from attacks, the
private key is often frequently changed, and the process of agreeing on a private key may need
to take place in person. Furthermore, increasing the number of users in this cryptosystem
also increases the chances that the system will be broken. Thus, private key cryptosystems
do not scale well. These difficulties are not present in public key cryptography.
    \subsubsection{Asymmetric Cryptography}\label{asym}
Public key cryptography, also known as asymmetric key cryptography, takes
another approach to the process of encrypting and decrypting. In public key cryptography each one, Alice and Bob, maintain their own distinct private key and also
a distinct public key. A public key is a piece of information that is published for all
parties to see. Thus, Bob and Alice each publish a public key but they also keep a
single piece of information secret. Only Alice knows her private key and only Bob
knows his private key. Note that an individuals public and private key are typically related in some way that facilitates and enables the encryption and decryption process.\par
\begin{figure}[htp]
    \centering
    \includegraphics[width=0.7\textwidth]{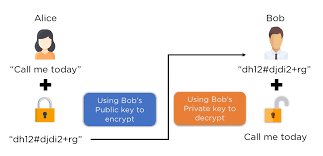}
    \caption{Asymmetric (or public key) cryptography}
    \label{fig:public_key}
\end{figure}

Suppose Alice wishes to send Bob a message (fig.\ref{fig:public_key}). Alice begins by looking up Bob’s public key. Alice then uses this public key to encrypt her message and transmits
the result to Bob. Bob receives Alice’s transmitted message and uses his secret
private key to decrypt the encrypted message and recover Alice’s original message.
Notice that the information available to an attacker has increased substantially. The
attacker now has knowledge of a public key (which is related to the private key used
for decryption), additionally to the ciphertexts. In order to ensure security, it must
therefore be diﬃcult to produce the private key from the public key. A well-known
implementation of public key cryptography, RSA.

\subsubsection{Hash Functions}
\begin{figure}[htp]
    \centering
    \includegraphics[width=0.7\textwidth]{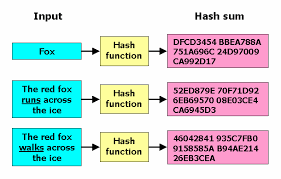}
    \caption{Hash function produces same length output }
    \label{fig:hash}
\end{figure}

This type of encryption doesn’t make use of keys. It uses a cipher to generate a hash value of a fixed length from the plaintext. It is nearly impossible for the contents of plain text to be recovered from the ciphertext. Therefore, the hash function is a unique identifier for any given piece of content. In this process, plaintext data of any size is converted into a unique ciphertext of a specific length (fig. \ref{fig:hash}).\par
By looking at the definition of hash function it may appear very similar to encryption yet hashing and encryption are not the same. The very basic difference between the two is that, unlike encryption, hashing function does not require anything like decrypting the hash value. It basically works in a way that plaintext data is inserted and using a mathematical algorithm an unreadable output is generated. The output is called hash digest, hash value or hash code, which is the unique identifier. Properties of a strong hash algorithm include determinism, pre-image resistance, collision resistance, good speed and avalanche effect aka snowball effect. 
Hash functions are the key element of a wide and well known technology, the blockchain.
\subsection{The Blockchain}\label{ch:blockchain}
Blockchain is a chain of blocks where each block contains a set of transactions that are digitally signed by its "verifier" and stored across the distributed network so that all the legitimate stakeholders can access/verify them\cite{blockchain}. Due to the attributes of Blockchain such as decentralization, immutability, auditability, transparency, and cryptographic security, it offers various benefits to different domains such as cryptocurrency, financial sectors, private/public segments, insurance, healthcare, supply chain management, Internet of Things, etc. Blockchains are typically managed by a peer-to-peer (P2P) computer network for use as a public distributed ledger, where nodes collectively adhere to a consensus algorithm protocol to add and validate new transaction blocks. Blockchain records are not unalterable,  considered secure by design, with high fault tolerance.\par
 Blockchain technology provides various benefits as follow:
 \begin{itemize}
     \item \textit{Transparency}: Transactions stored on the Blockchain are transparent to all the participated users. Blockchain uses the distributed ledger (a shared copy of document) kept by individual parties and can only be updated by the consensus mechanism, which means that the file can only be updated if all the legitimate parties agree to do so.
     \item \textit{Security}: There are many ways by which blockchain is more secure than the other record management systems. Transactions are added after the consensus by all permitted parties. Once everyone agrees upon the transaction, it is encrypted and securely linked with the previous block. Secured hashing mechanisms attached with each block are used to secure the blocks that hold the number of transactions. And hence, it is practically infeasible to temper a block as it requires modifications to other blocks in the chain too.
     \item \textit{Traceability}: Tracking of data/process is easy with Blockchain. Transactions are visible to all parties which lead to traceability for any operation. If enterprise deals with the supply chain, the tracking of the product is easy through this technology.
     \item \textit{Fast and Efficient}: In a traditional system, the paperwork is time-consuming, tedious, and prone to human errors. By automating it with Blockchain, the process becomes more fast and efficient and operates without any third-party intervention.
     \item \textit{Cost-effective}: For any business, profit/cost-effectiveness is important. With this technology, it doesn’t need any intermediary or third party. Hence, it becomes cost-effective.
\end{itemize}
 Blockchain technology became popular and known after it was introduced and used in cryptocurrencies, such as Bitcoin, introduced by Nakamoto, in 2008. Bitcoin was the first electronic payment system without third-party intervention using decentralized and distributed peer-to-peer networks. The term “Block” and “chain” used separately by Satoshi Nakamoto.\par
\begin{figure}
    \centering
    \includegraphics[width=0.75\linewidth]{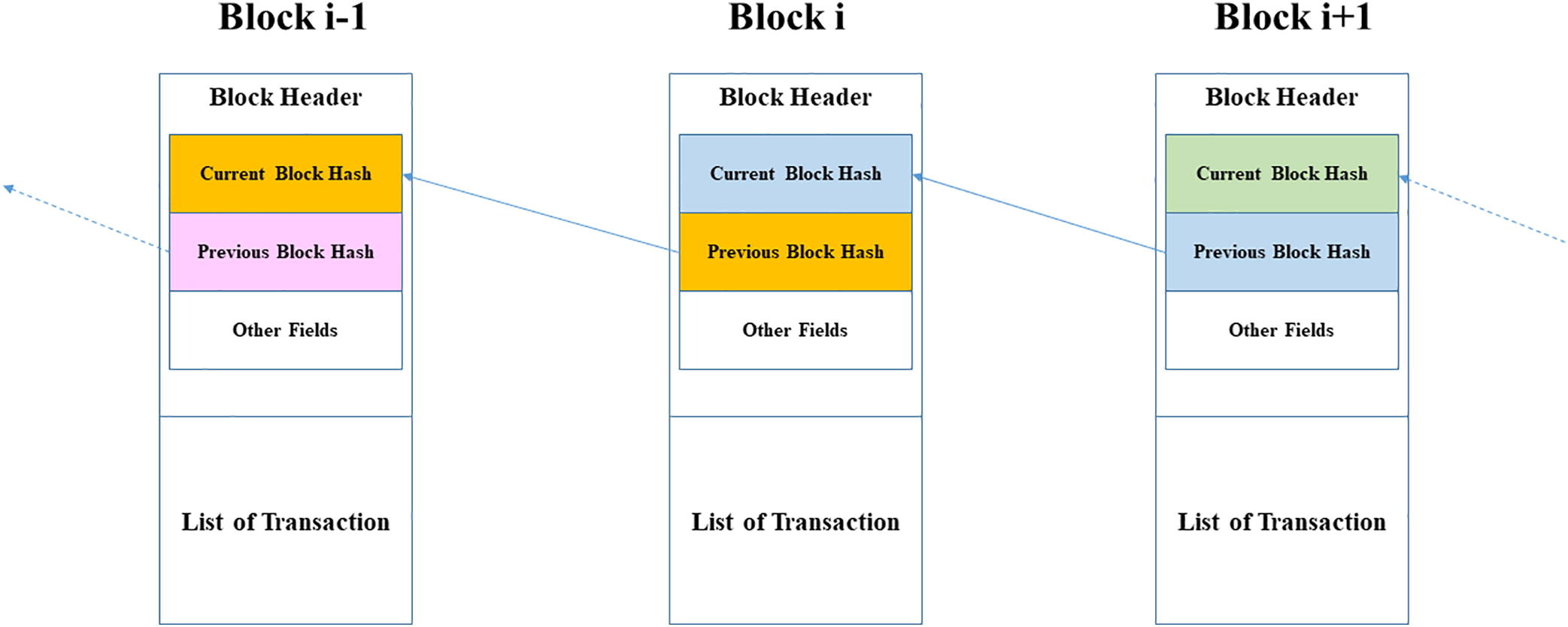}
    \caption{Blockchain architecture}
    \label{fig:chain_arch}
\end{figure}

  Blockchain, in a simple word, is a technology that provides accessible and verifiable data control over the distributed (decentralized) environment to every participated node in a fast and convenient way. There is no single or centralized authority to validate/verify the nodes. In order to  participate in a network, a node has to validate itself by solving a mathematical puzzle called a proof of work. A node that succeeds in a proof of work can introduce a block. We will see the block and its content in detail in the next subsection called architecture. The action that  new data that must be validated and become content of a blockchain is initiated by a node and call \textit{transaction}.
Blockchain's architecture can be seen in fig.\ref{fig:chain_arch}. 

Then, a blockchain’s subsequents and subparts  are:
\begin{itemize}
    \item \textit{block}:  chunk of data (or a few transactions) is grouped to form a block in a blockchain
    \item \textit{transaction}: a chunk of data to be stored on the blockchain
    \item \textit{transaction verification sequence}:A user generates a transaction and sends it to all the nodes on the network, then
Nodes verify this transaction and add it to their candidate blocks, then nodes then broadcast their candidate blocks to the entire network, and so on and so forth.
\end{itemize}\par 
 Blockchains are separated in two types. The \textit{public}, which are less safe due to lack of restrictions implemented and simultaneously slow because they  open to anyone who want to become node. On the other hand, there are the private blockchains that are fast and suppose to be safe. Safety though is a matter of assumptions we make because whenever a provider is private there are few ways to investigate its mechanisms. 

\section{Unmanned Aircraft System Traffic Management (UTM)}
\subsection{General}
The aforementioned chapters \ref{iot}, \ref{uav} generate the strong belief that  in the near future our skies will occupied by several kinds of aerial vehicles. This prediction sets as mandatory to  reconfigure the control of the airspace. Until now, the Ground Control Systems  encapsulate all the modules and functions under rules related to classic aviation, which are the manned vehicles. Moreover, the manned vehicles per se, include equipment capable to interchange data with specific format and under specific variety of other vehicles. For example, manned aerial vehicles utilize Radio Frequencies in order to communicate or navigate towards their destination. Moreover, due to the future congestion of the airspace, the communication and guidance infrastructures must manage the diversity consists of   aerial vehicles. Thus, there is a growing research towards the   Unmanned Aircraft System Traffic Management (UTM).\par
A UTM is designed to facilitate the integration of the UAVs into the National Airspace System (NAS), particularly in the Very Low Level (VLL) airspace, with utmost emphasis on safety and security. The VLL airspace is set to under of 400 feet Above Ground Level (AGL). To achieve this objective, the UTM provides distinct services to the stakeholders that comply with respective country regulations and separate from those offered by ATM. The physical integration of the UAVs in the NAS entails collaborative efforts between the ATM and UTM, encompassing both regulatory and technological endeavors. Some of the fields that a UTM focuses are the following:
\begin{itemize}
    \item Airspace design and operation
    \item Physical and communication infrastructures
    \item Technical and communication standards, protocols
    \item Regulations
\end{itemize}
A UTM shall be costructed by several agents. For instance, if there is one supplier that provides UTM services, the UTM is classified a monolithic. On the other hand, if there is more than one is called federated.

\subsection{UTM Architectures}
In order to to use the airspace with both manned and unmanned vehicles we have to introduce an organization that services the stakeholders who want to utilize the airspace. This  organizations are called UTMs and they are divided in two main categories based on their architecture, the centralized and decentralised UTMs. Both of them though, share the same principles, must offer the same functions and includes almost the same entities. Some of these entities are:
\begin{itemize}
    \item The UAV operator
    \item The UAV itself
    \item A UAS Service that allow operators to use the UTM
    \item Supplemental Data Service Providers (SDSPs) that provide additional information and data, like weather conditions of terrain morphology. 
    \item A regulator that is responsible to authorize all the parts and make them capable to be part of this ecosystem. The major aim of the regulator is safety.
\end{itemize}

The key element that distinguish discrepancies are the way the data in exchanged between the stakeholder. Furthermore, if there is one entity that controls the data exchange then this kind of architecture is considered as \textit{centralized}. On the other hand, if the stakeholders that utilize the UTM are free to exchange data in both open or closed manner, without a central entity that controls data flow, then this kind of architecture is consider \textit{decentralized}.

\subsection{Centralized Architecture}
A centralized UTM  is based on a central entity having the ability to support
relationships and information flow between the stakeholders. The foundation of a centralized
UTM’s architecture is argued by the safety issue. A single trustworthy source of critical information is mostly considered more accurate and safer. While a centralized system in considered to be more affordable, they have an significant drawback related to the theory of the \textit{single} \textit{point} \textit{failure}\cite{single_p_f}. There are a variety of factors that may trigger a single point failure incident which jeopardises the safety of the centralized UTM. Some of these factors may be:
\begin{itemize}
    \item Pilot errors
    \item UAV technical malfunctions
    \item Bad weather conditions
    \item Unexpected events, like drop of the power supply       
\end{itemize}

As the number of the UAVs increases the airspace is getting more and more dense. It is an analogy of the command and managing principles, where the more people must be managed and more levels of management must be introduced. Thus, when a lot of aerial vehicles fly on the sky, the possibility to take place an incident based on the aforementioned factors is higher. For instance, if two or more UAVs declare an emergency simultaneously, then the centralized entity must manage this unwanted situation. The dilemma that arises then is where to  provide the the most of the sources. If the UTM spent the bandwidth and the computational capability to handled the emergency, then the other vehicles that are still on the sky may encounter high latency. Hence, the possibility to have  the sequence of possible accidents increases.\par
In other words, the aforementioned latency, combined with sudden changes in velocity or trajectory, can negatively impact the efficiency of the decision-making process. Therefore, the combination of the dynamic behavior of UAVs, the increasing complexity of the airspace with a growing number of UAVs, and the latency introduced by centralized communication poses significant challenges to the effectiveness and efficiency of a centralized UTM system.\par
\begin{figure}
    \centering
    \includegraphics[width=0.75\linewidth]{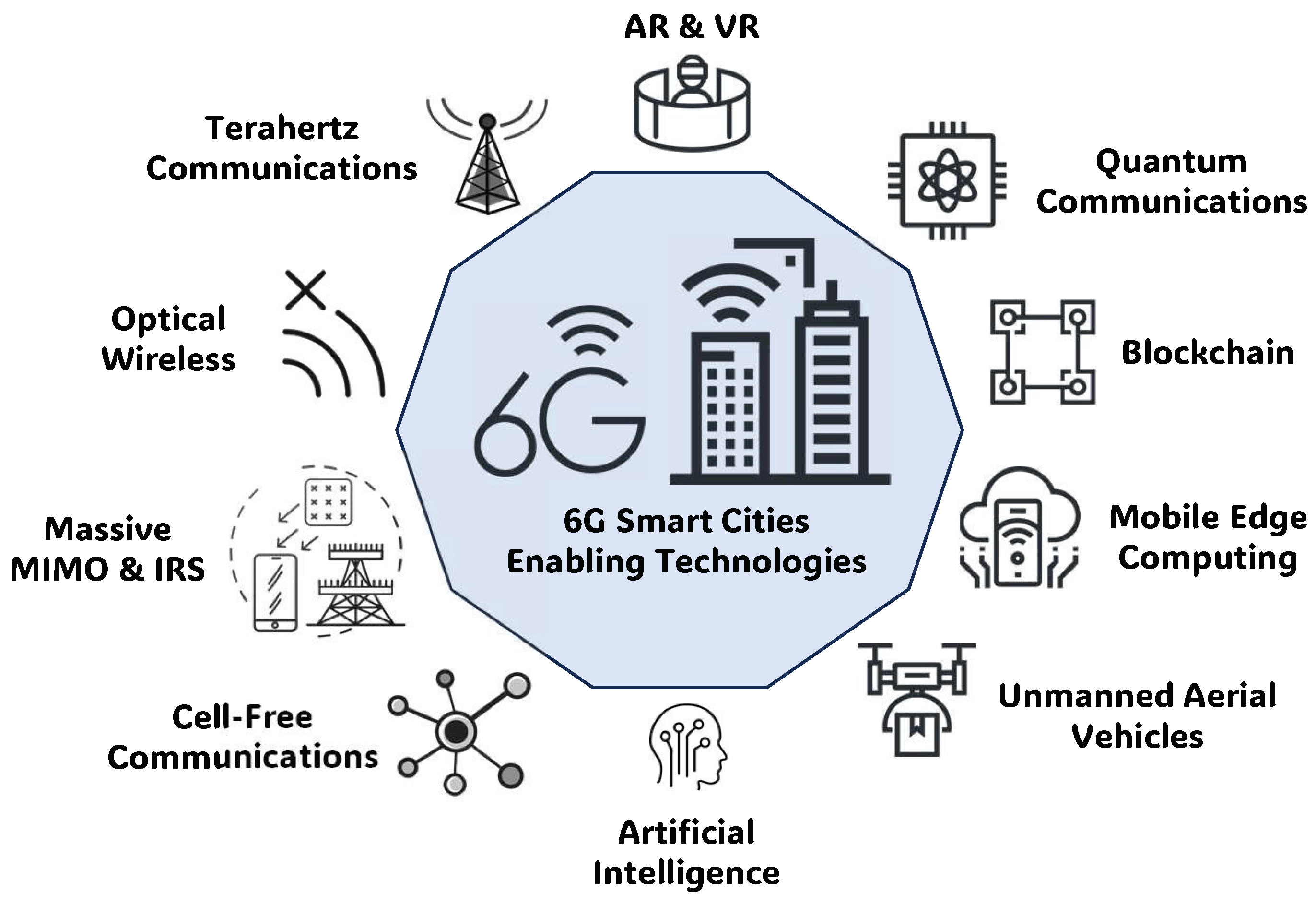}
    \caption{Usage of 6G in future.}
    \label{fig:6g}
\end{figure}
\begin{figure}
    \centering
    \includegraphics[width=0.75\linewidth]{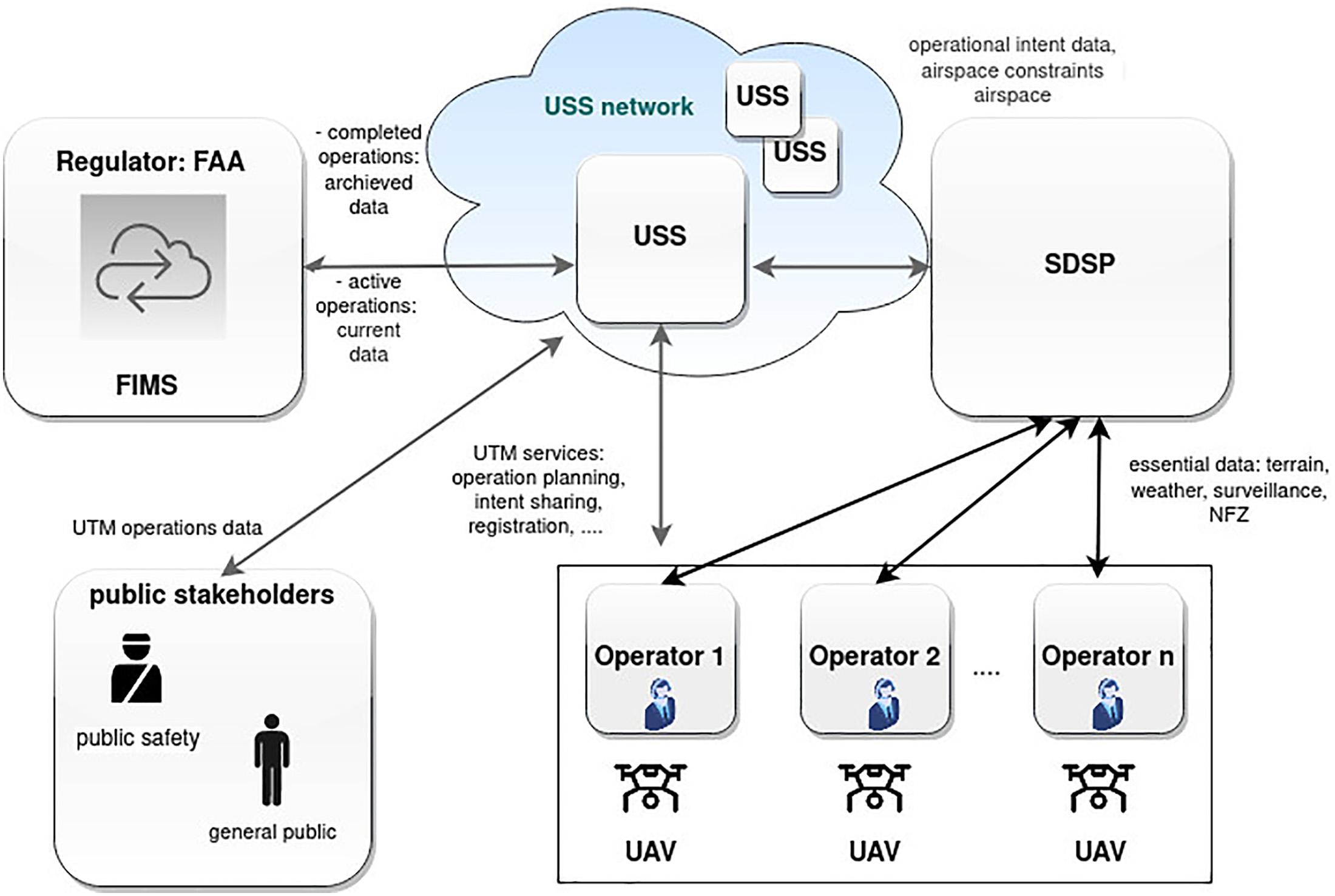}
    \caption{US UTM architecture}
    \label{fig:us_utm}
\end{figure}
\begin{figure}
    \centering
    \includegraphics[width=0.75\linewidth]{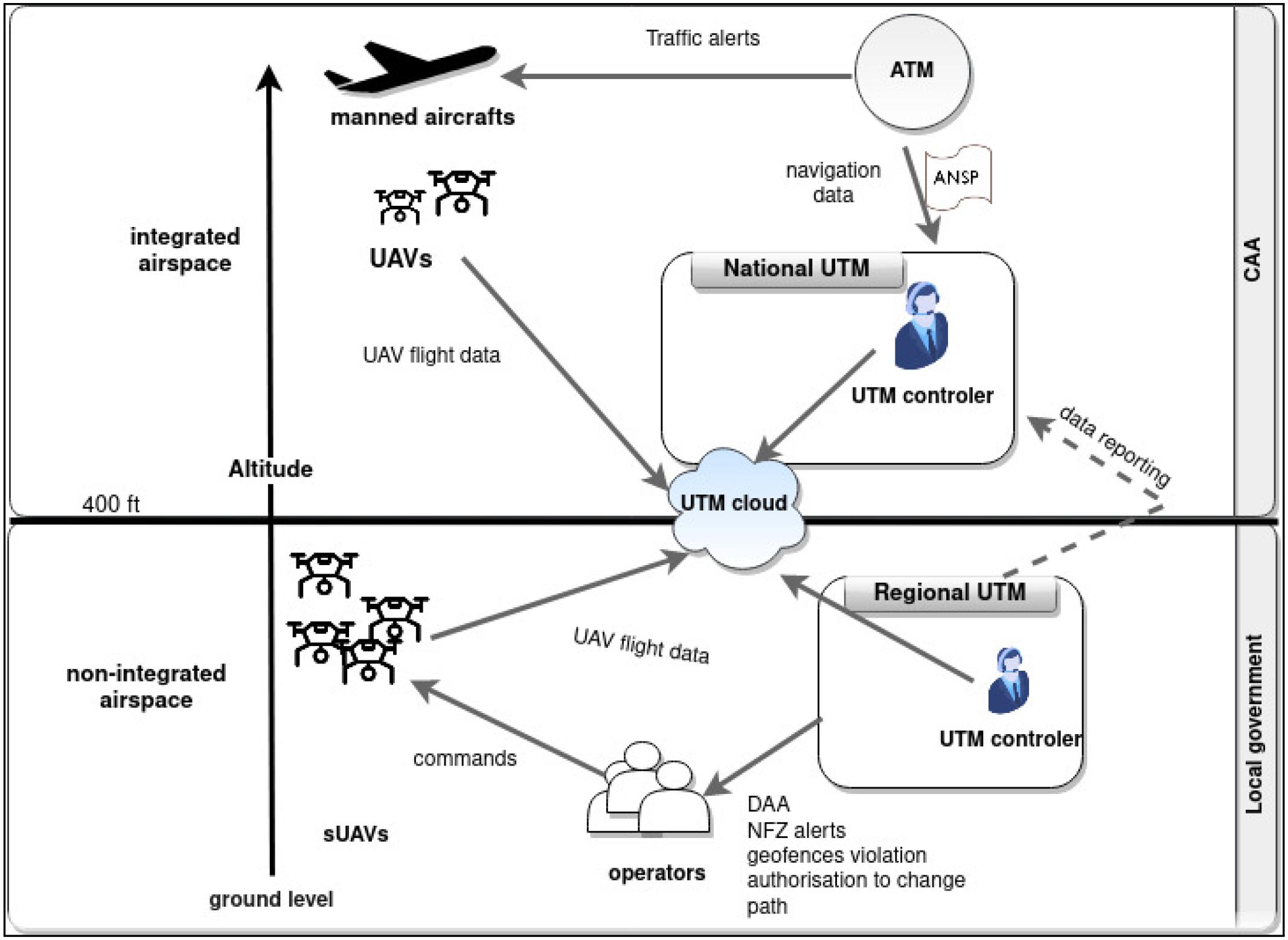}
    \caption{Hierarchical Centralized Architecture}
    \label{fig:hier_utm}
\end{figure}

Considering the criteria that establish an architecture as centralized, we may distinguish these entities that one of them is in charge of managing the function of the UTM. Thus, we classify the centralized in those:
\begin{enumerate}
    \item \textit{Based on Regulator Rules.} A centralized architecture based on regulator rules relies on a central entity like the FAA (see ch. \ref{ch:Regulators}). This kind of centralized architecture seems to be significant strict. The rules that the operators must follow in order to utilize the UTM are co-related with the capabilities and the technical characteristics of the unmanned vehicles. Based on literature, there are two such architectures, the one from USA (US UTM) and the other from India. The former developed under the collaboration between the FAA and the National Aeronautics and Space Administration (NASA). This kind of architecture face issues of latency and scalability and thus the decision making procedures in case of emergency increments the overall latency, provoking additional risks.\par
US UTM backbone is based on constant communication between the regulator FAA (see about regulator in  ch. \ref{ch:FAA}) and the US UTM as shown in fig. \ref{fig:us_utm}.
    \item \textit{Hierarchical Centralized Architecture.} The rational behind the function of the Hierarchical Centralized Architecture is that the aerial vehicles to be separated according to the altitude. Thus, the limit set is 400 meters, something that remind us considering the VVL that contain the condensed  and crowded airspace from several vehicles. Taiwan is the country that implements this kind of architecture. In this case, the centralized entity that communicates with the stakeholders still exists, see fig. \ref{fig:hier_utm}. The centralized architecture includes the UTM cloud and a principal UTM server. The role of the UTM cloud is to receive UAV surveillance data, whereas the principal UTM server is to receive and treat data from the cloud. This approach focuses primarily on the surveillance function of the UTM.
    \item \textit{Centralized Service-oriented Architecture.} In this aspect of function, the UTM  reorganising the software services and infrastructure in a manner that they interact each other.  The architecture is simple and bases on integration,  sharing and the reuse of services from several providers. Thus, the services must use common interface standards and protocols to be rapidly incorporated into new applications.
    
    \item \textit{Centralized Architectures based on Specific Cellular Network.} Whereas cellular technologies  used in UTMs widely, some centralized architectures base their function on the scalability of the cellular concept. Cellular networks are based on high frequency radio waves.  Until now the technology  have been  introduced is 4G and  the corresponding frequencies are varied, including the 600 MHz, 700 MHz, 1700/2100 MHz, 2300 MHz, and 2500 MHz bands. The lower frequencies made it possible for carriers to transmit 4G/LTE signals in  remote areas. High frequency signals  need a mean to rely  signals (i.e satellites) to the receivers which are Beyond Visual Line Of Site (BVLOS). This happens  because this kind of signals used to be absorbed from mountains , buildings, etc.\par
    We are indeed facing an evolution regarding the telecommunications. The next era of such technologies is 6G  \cite{6g}, fig. \ref{fig:6g}. These networks will  be significant faster than previous generations and also expected to be more diverse. This generation will support applications beyond current mobile use scenarios. Concerning the very low  latency, the 6G generation will be  much more reliable than the upcoming 5G. This concept  implies that the communication systems that utilizing 6G shall be the next generation of the UTMs.  
    
\end{enumerate}


\subsection{Decentralized Architecture}\label{sec:dec-arch}
In the previous chapter mentioned that the main drawback of the centralized systems is the \textit{single point failure}. This  flaw can be extinguished by setting the concept of the blockchain as the backbone of a UTM. This framework is named \textit{Decentralized (or Distributed) Architecture}. Some of the big companies in the aviation industry founded research for the development of UTMs based totally on a decentralized architecture. In ch. \ref{ch:blockchain} analyzed the  fundamentals of a blockchain's architecture where security is one of the key element, protecting UTMs from attacks like signal spoofing, man in the middle etc.\par
\begin{figure}
     \centering
     \includegraphics[width=0.75\linewidth]{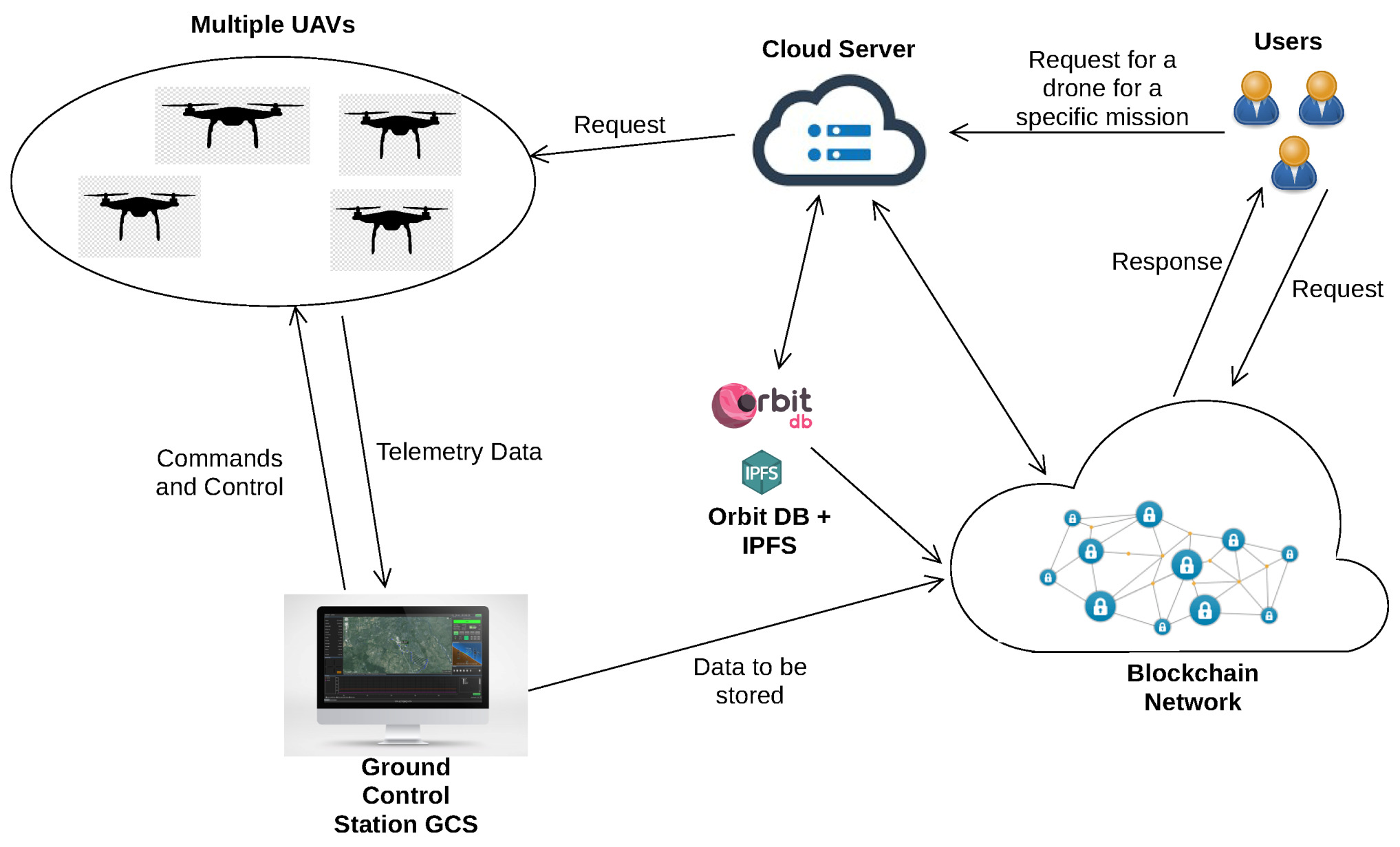}
     \caption{Blockchain UTM architecture}T
and the states of the smart contracts.
     \label{fig:block_UTM}
 \end{figure}

In order to understand the decentralized concept, there are several entities that participate into this ecosystem, which are:
\begin{itemize}
    \item UAVs as autonomous or remote piloted vehicles
    \item Ground Control Stations (GCS) which will guide the drones, control the airspace and ensure collision avoidance. The GCS receive data from the drones and send to them Command and Control signals (C2)
    \item The blockchain network. This will serve distributed database for sharing recorded transactions among network nodes
    \item Cloud server to assist drones' computations
    \item Users that use data from UAVs and GCS from several reasons.
\end{itemize}
 All the aforementioned entities  are solving the problem of the safe and uninterrupted data manipulation but not storing them. Hence, there is another database called ORBIT DB (see fig. \ref{fig:block_UTM} introduced to ensure that all the data will be available. The data stored are about the drone, its operator and its mission. \par
 In order to accomplish a mission, each drone and user must have a unique Remote ID or RID(see ch. \ref{ch:FAA}). Firstly, the operator must register its UAV to be assigned an ID and be added in the authority database which shall broadcast during the flight. Then, he must subscribe the flight plan in order to request the services of the UTM.\par

 The objective of the conflict management is to avoid physical collisions between the UAVs and static or mobile obstacles namely the buildings and structures, people, animals, manned aircraft and other UAVs. There are studies  present  conflict management algorithms as a process named  strategic deconfliction,  tactical deconfliction. The strategic deconfliction (mission scheduling) is done before the UAV flight. To ensure collision avoidance, the operators to submit their flight intent based on precise information about weather, terrain, and airspace constraints. Despite the fact that submitting a flight plan, the risk of collision is still high. Thus,  decision-making is a concept that provides the solution to be adopted and is attributed to the emergency management procedures. When multiple UAV flight plans conflict due to a loss of separation, the solution is to modify flight plans by using a geometric approach. As the principle of the blockchain is that the data are pert of all nodes, the new flight plan generated is instantly submitted as a transaction.

\section{Communications in a UTM}
Currently, the pilot and the ground control tower mainly communicate via UHF (Ultra High Frequency, around 1 GHz ) or VHF (Very High Frequency, around 100-300MHz) analog voice radios ~\cite{analogtodigital}.When an analog voice radio communication technology is used, all pilots in the same sector in order to communicate with an air traffic controller must be  tuned to the same frequency. This can be challenging considering the expected air traffic growth. Statistical data on air traffic reveals that there is an  increase  trend regarding the  transportation industry over the air. Long term forecast studies provided by Boeing predict a \textit{5\%} growth rate of the world air traffic load between 2011 and 2030 . This growth is due to many factors such as the  more competitive low-cost airlines, the  increased passenger demand to travel and the greater need for companies to provide a better service to their customers. Nowadays, the air traffic load is still increasing, leading to a congestion of the worldwide analog voice frequency allocated to the civil aviation.

The term "Data-Link"  is commonly used among the civil and military aviation community as the digital communications between an aircrafts and a ground stations (A2G) or between aircrafts to aircrafts (A2A). There are several benefits that digital data exchange is preferred. For instance, using digital way to transmit signals we are able to:
\begin{enumerate}
    \item Confirm the ground instructions aircrafts receive via special on board devices
    \item Correct errors during transmission. Thus, no data loss due to signal jamming.
\end{enumerate}
The UTM is based on information flow exchange between all participating entities, making the
communication links crucial to conduct command and collision avoidance functions. The needs
in communication links differ according to the level of autonomy and the extent between the
operator and the UAV, for both Visual Line Of Site (VLOS) and Beyond Visual Line Of Site (BVLOS) cases. The communication technologies used for UTMs must guarantee a real-time communication, a wide coverage area, a secure networking and a very low latency. The evolution of the cellular communications \cite{6g}  may contribute to lower the latency and establish the UTMs communication to tend in a real-time fashion.

\subsection{Command and Control (C2) Communication}
The Command and Control (C2) communication ensures that, after being registered, a UAV is constantly under control even if it has to modify its mission or its trajectory. The controller may be a GCS or a dedicated controller. Hence, the C2 concept must be ether between GCS and UAV or between the controller and UAV and the Ground Radio Station and must ensure that this data will  be transmitted without any errors. For UAVs there are three flight modes: lateral flight, vertical flight and hovering, with the possibility to change the mode.The information transmitted in a control message differ according to the mode applied. Specific values of key performance indicators (KPI) are associated with each C2 mode.\par

Several technologies are used to ensure C2 communication. They can vary from terrestrial cellular networks to satellite connectivity. In fact, if the operation is VLOS, the communication is done through a direct radio link. A BVLOS communication relies on the use of a satellite connection \cite{C2} or on more than one connection with redundant C2 links. For the decentralized blockchain UTM systems, the GCS receives data from the UAV and then returns the control
commands using the cloud through the Internet.  While the UAVs are connected to the Internet, the data security is in great hazard because is  exposed to malicious attacks. Threats can vary  and some of them are referred in ch. \ref{ch:Data_sec}, \ref{ch:Data_sec_quantum}. Their main purpose is to alter the initial data stream reducing  the stability and security of the UTM.

\subsection{Air-to-Air (A2A) Communication}
In ch. \ref{uav} was implied that there is an  incremental increase of the air traffic, foreseeing that  the next decades air traffic management will come to a totally new era. Specifically, new air traffic services and operational concepts have been defined and shall be supported during all phases of flight. A set of modern digital data links integrated into a single communications network. \par
LDACS is a cell-based aeronautical communications system operating around 1000 MHz  frequency band and  supports data and voice communications between ground stations and airborne stations \cite{a2a_ldacs_mode}. LDACS is a wideband terrestrial system with VLOS coverage intended to work along VHF Data-Link2 (VDL2) for new and more demanding services. It operates in the L-band (around 1 GHz), which has excellent propagation characteristics. The operational compatibility (spectrum interference) with existing L-band systems, such as  navigation Distance Measurement Equipment (DME), Global Navigation Satellite System (GNSS), Military LINK16 and mobile telephony, remains an important subject, and desirable technology features were identified that could make LDACS spectrally compatible.
An LDACS ground station is located in the center of each cell and communicates with the LDACS airborne stations located in the aircraft flying within its cell. Using a frequency-division duplex scheme, the LDACS ground station transmits in the forward link (FL) of the cell at the same time as the airborne stations transmit in its reverse link (RL). In 2019 took place the first experimental usage of the LDACS \cite{ldacs}.  In this experimental  four simulated ground stations (GS) created as an airborne and an aircraft occupied for the test flight. During the  LDACS experiment,  six flights took part and  those produced numerous  data, addressing the communication, navigation and surveillance capabilities of LDACS. Furthermore, there implementation consists of an Ground Control Station (GCS).\par

The air-to-air (A2A) communication within the L-band Digital Aeronautical Communications System (LDACS) is currently in the initial stages of its development \cite{a2a_ldacs_mode}. Given that, the LDACS A2A mode must be able to operate without any ground or satellite support, the data link must provide means for the aircraft to establish and organize an independent communications ad-hoc network, which imposes a great challenge for the design of the data link and specially for its medium-access control.\par

The Air-to-Air (A2A) links concern both the UAV-to-manned aircraft and UAV-to-UAV (U2U)
communications. The UAV and the manned aircraft must exchange positional information to avoid
collision, especially when the UAV is inside the controlled airspace or around the airports. The radio technology is used to allow the two-way radio communication with Air Traffic Control (ATC). The U2U communication is especially used in the decentralized UTM to exchange
the essential data necessary for the local decision-making in collision avoidance. Besides that cellular communication is proposed to be used in the decentralized architectures \cite{C2_cell}, Wi-Fi and Bluetooth protocols may be used in case of conflict management in short range distances.\par

\section{Decision Making in UTM}\label{ch:decisnion_make}
During execution its flight plan, a UAV may be  exposed to external hazards or face an internal malfunction. The level of risk associated with the UAV can vary, resulting in varying degrees of impact on its immediate environment. In some cases, this could lead to incidents or accidents and relating the distance from the airport, an accident may be potential deadly. Therefore, effective decision-making within UTM systems is crucial, especially in  emergency situations.\par
In general, an emergency situation may occur for  several reasons, such as:
\begin{itemize}
    \item Technical failures, such as the loss of A2G communications, Global Positioning
System (GPS) malfunctions, power, camera, or engine failures.
    \item Human errors, involving control or wrong decision making.
    \item Corrupted data due to cyber attacks
    \item Infrastructure problems such as radio control failures
    \item Sudden weather condition changes
\end{itemize}

In case of one of the above parameters occurs, the UAV has the option to either halt the mission  or alter its flight path. Onboard decision-making procedure are encapsulated to minimize 
air-to-ground dependencies. A Finite State Machine (FSM) is implemented, in order to  improve the quality of decision making  response time. On the other hand, there is another rational that manipulates predicts risk factors and  assesses the decision making based on these calculations. These kind of factors may be generated from the reliability of the UAV's performance. \par

The decision is triggered  by a combination  of the \textit{risk-factor prediction}, the \textit{decision generator} and \textit{trajectory generator}. The three functions can be implemented totally on-board the vehicle allowing UAV to be independent concerning communications. In fact, on board decision-making process serves
to effectively reduce the necessity for ongoing data transmission between the UTM and UAVs,
a factor of considerable significance, particularly in scenarios involving communication disruptions. However, it requires an on-board computational capacity. An alternative architecture involves conducting the risk prediction remotely, while maintaining the other two functions on-board. It allows the
resource sharing and a reduced size and weight on-board but the decision-making is dependent on
a communication that must be robust and unceasing. This architecture is based on remotely risk prediction and path generation with on-board decision-making. In the final architecture all the functions are hosted remotely. Here, the decision-making is widely dependent on the network connection.
\subsection{Centralized vs Decentralized UTM Decision-Making}

According to \cite{message_categ}, the messages that a UAV exchange are categorised to:
\begin{itemize}
    \item \textit{Control messages}: are sent from the operators to the UAV.
    \item \textit{Telemetry messages}: include the Global Navigation Satellite System (GNSS) position, status, and altitude information.
    \item \textit{Awareness messages}: contain information about the current position of the UAV and its future
trajectory.
\end{itemize}
A UAV is supposed to be autonomous when it is able to accomplish its mission without any input from the 
operator. Thus, the more autonomous a vehicle is the less telemetry and awareness signals need to transmit. A high level of autonomy means a capacity of self-decision making and the possibility of communication between UAVs  in order to avoid conflicts. For military UAVs, the automation levels must be linked to detection and tracking capabilities and a self tactical deconfliction aptitude. As mentioned in the previous paragraph, this affect vehicles's mass.\par

In the centralized architecture, if a conflict is detected, the information flow from the UTM to the operator contains the updated trajectories, which in turn, transmits the new commands to the UAV. All these data are transmitted through the same channel shared with  the other aerial vehicles. The latter use this channel to execute their flight plan and consequently this  increases the latency. Moreover, apart from receiving all these data, the centralized entity ought to  process and update new date not only to the emergency part but to the other vehicles which leads to a higher demand of quality of service, sources, etc. In this state of function, the decisions made are not representative against the real situation that the system is. The latency and the processing time referred to the data received some ms before and do not represent the current status.  \par

The case is different in a distributed decision-making (decentralized) system. The awareness data is no longer transmitted to the UTM but is exchanged between the UAVs. If a conflict is detected, the concerned UAV makes a self decision or reacts by relying on a robust and aligned mechanism. Conflict resolution is a process based on the communication between the UAVs more than the A2G communication. The direct communication between the UAVs and the local decision-making reduces the decision latency. In this decentralized architecure, the degree of collaboration between UAVs implies the existence of three categories: 

\begin{itemize}
    \item \textit{Decentralized with uniform rules}. Each UAV resolves conflicts by
executing a set of predefined rules and protocols. The communication between the UAVs is only
to share necessary data.
    \item \textit{Decentralized with coordination}. The UAVs get new coordination according to preset manner.
    \item \textit{Decentralized with mixed rules}. Depending on the area they fly, UAVs follow a certain batch of rules.
\end{itemize}

\section{Performance Evaluation of a UTM }

A UTM is a perplexing ecosystem consisted of subsystems where the total performance is the sum of the performance of each part. The communication-navigation and the decision-making policy are major subsystems that can been evaluated separately. NASA proposed some quality and quantity metrics that a can evaluate the aforementioned performance. These metrics are values that asses the UTM  in emergency cases, conflicts, and communication loss to study the system reaction in such situations. The performance is measured based on resource usage
(memory usage, CPU consumption) and the network latency, which is the total time duration taken
for the execution of a transaction in the blockchain network.\par

In order apply all these metrics we must built such a system and furthermore to compare the results with other UTMs. This case seems to be difficult to be monitored. Hence, it is essential to apply  simulations in a UTM where the input may be some critical situations such us these mentioned in ch. \ref{ch:decisnion_make}.\par

In \cite{u-space}, a simulation framework based on the Robot Operating System (ROS) and the simulator Gazebo. This simulator uses the architecture of U-space. U-space is a set of specific services and procedures designed to ensure safe and efficient access to airspace for a large number of drones, and which are based on high levels of digitalisation and automation. Using ROS, the UTM is modeled as a set of independent nodes. The UTM manager node reflects the aspect of the system, and the DB manager is the node linked to all databases and responsible for the basic write and read operations. The main pre-flight and
in-flight services namely the flight plan management, tracking, monitoring, emergency management, and conflict solver of U-space are modeled as nodes too. \par
UTSim \cite{utsim} is another simulator that is able to evaluate the performance of a UTM. Its primary advantage is that can use 3D models and represent  the nodes of the UTM in a more fancy way. Another important  is the scalability and collision scenario
in which the number of UAVs that are allowed to fly simultaneously was varied from 20 to 1500. Firstly, the simulate a system that include drones to fly without implementeting a deconfliction algorithm. Then, by inserting several deconfliction algorithms to system, the concluded that the number of collisions is linearly proportional to the number of UAVs flying. The deconfliction is tested
with the Barfield’s algorithm \cite{Barfield_2020}, which is an existing collision avoidance algorithm based on geometric method. Additionally, it is crucial  to test the simulator with more deconflicting algorithms and to insert the  deconfliction in the integrated airspace between manned and unmanned aircraft.

\section{UTMs Open Issues}
In nowadays, the UTMs are in heavy development and a lot work must be done in order to be ready to manage the airspace. The era we are living is an evolutionary era where the science of communication is beyond all in doubt. Techniques  that used to be fundamentals in communications are  about to considered obsolete. On the other hand, new technologies are about to extinguish the human factor which is considered  from some people controversial.
\subsection{Interoperability}
\textit{Interoperability} is an internal function where all the different parts of the UTM ecosystem must integrate each other with harmony. In case of an entity does not set a communication channel between another one, then there is not a smooth interoperability.\par
In the centralized architecture, the interoperability is a fact due to the reduced parts that must exchange data. On the contrary, in the distributed systems the entities differ. Hence, there must be introduced compatible protocols, data format and standards for flawless communication. In other words, standardisation not only must be introduced in any aspect of the UTMs but also among all the  UTMs. Fortunately, there are several agents and authorities that manages standard (see ch. \ref{ch:Regulators}) which are in a high state of consciousness and aware of the next era in aviation and airspace management. 

\subsection{AI Issues}

FAA and EASA have initiated discussion around Artificial Intelligence (AI). Generally speaking everyone has a different definition of what AI is. AI allows machines to learn from experience and adjust the way they respond based on the new data they collect. In other words, AI goes through a learning procedure in order to be able to act based on its "thinking". Traditional aviation software is certified to be Deterministic via guidelines such as DO-178C (avionics software) \cite{Do-178c} and DO-254 (Avionics Hardware) \cite{do-254}. AI essentially enables the same software inputs to yield a different outcome as the software "learns" over time. The main concern around implementing AI into transportation services is safety. Many entities, including the FAA and Department of Defense, look at AI through a "guilty until proven innocent" lens. One fundamental aspect of safety-critical systems is determinism, almost opposing AI, where the same inputs provide the same outputs, every time. This is where DO-178C comes into play. DO-178C is a set of guidelines covering 71 Objectives to ensure that software will perform safely in an airborne environment. The guidelines categorize software on five levels of Reliability, ranging from "No Safety Effect" to "Catastrophic". \par

Artificial intelligence will necessitate high levels of automation and act as an enabler with respect to the integration of unmanned and manned aviation and will ultimately enable safe operations with respect to high numbers of drones utilising the same airspace, and more specifically with respect to detect and avoid capability. AI is going to be heavily developed and utilised by organisations that certify as U-space service providers (USSP’s) when providing a service to Unmanned Aerial Systems (UAS) operators. The equipment utilised by UAS operators will to some extent already benefit from AI but the level of automation is currently constrained by regulation. A legal framework must exist as AI will not only have a significant impact upon existing laws but will ensure a framework that facilitates safety and the fundamental rights of citizens and businesses, with respect to AI. The EU has published a proposed law, namely the Artificial Intelligence Act as permitted under Article 114 of the Treaty on the Functioning of the European Union (TFEU) \cite{ai_in_UTM}.

In a UTM, AI focuses in collision avoidance, detecting and preventing potential conflicts between UAVs and other objects such as buildings, aircraft, or other UAVs. An issue that may consider is how much  time AI algorithms need to evaluate and learn from new data. As a matter of fact, artificial intelligence  systems that make decisions based on historical data. Thus, AI must learn in a simulation mode, initially. While the concept of simulating UTM is simultaneously in recent stages, then the implementation of AI in this period of time affects the evolution of the concept per se.  \par

Furthermore,  research has shown that artificial intelligence (AI) systems often include bias against minority subgroups \cite{ai_bias}. How this drawback may affect decision-making in a UTMs ecosystem? Regarding this assumption, the regulators must be aware of these algorithms and more precisely they must set the roadmap. Indeed, civil and military drones may share the same skies but considering their missions military drones seems to be more rock solid. This seems to be good but also seems to set military drones as the ultimate vehicle flying in skies. In fact, most AI-based systems are perceived as a black-box that allows powerful predictions, but it cannot be directly interpreted due to the difficulties in determining how and why it makes certain decisions. Actually, lack of transparency and trust in modern AI systems poses important ethical issues as highlighted in the "ethical guidelines" \cite{ai_attacks}.\par

In \cite{milit_drones_ai} proposed a way to mitigate such issues, implementing ethical rules. This approach based on quantitative ethics determines which action maximizes benefit and minimizes harm. Its objective is to make it possible for an AI algorithm to take the right decisions particularly when it encounters an ethical dilemma. \par

Additionally to the aforementioned internal flaws, in the  more technical view, machine learning and AI may suffer for several kind of outer parameters, like attacks on data. \cite{ai_attacks} mentions that blockchain and AI (two of the fundamental parts of the decentralized UTM architecture) have been recently found vulnerable to several cyberattacks and a number of security issues have arisen, especially when it comes to processing sensitive data. AI systems are also vulnerable to adversarial attacks, which become an inherent weakness of Machine Learning and  Deep Learning models.

\subsection{Data Security Against Classical Computers}\label{ch:Data_sec}
The information  while a  UAV transmits includes remote control commands, telemetry information, and mission sensor information. A remote control command is sent from the GCS to the targeted UAV. The main function is to control the UAV flight attitude and then guide it to the designated position and control the work of the mission equipment. Telemetry information includes aircraft attitude, flight parameters, equipment status and other related information that the UAV sends to the GCS. Regarding the remote control and telemetry information, these data sizes are very small. The  transmission is not high  but it requires real-time, reliable and secure transmission \cite{uav_comm_sec}. Mission sensor information refers to the information obtained by the UAV mission equipment, such as cameras, infrared scanners, multi-spectral sensors, synthetic aperture radar, etc. The data volume of each mission sensor node is related to factors such as sensor type, image format size, resolution, and data compression technique.\par

\subsubsection{Miscellaneous Attacks}
Communication security is crucial and  important for the success of Unmanned Aerial Vehicles (UAVs). With the increasing use of UAVs in military and civilian applications, they often carry sensitive information that eavesdropper would like to retrieve for several reasons. While UAVs consist of various hardware and software  modules, potential security vulnerabilities may also exist in those modules. For example, by launching a GPS spoofing attack or WiFi attack, eavesdroppers can capture the targeted UAV and access the sought after information.\par
Regardless the architecture that a UTM has been built, the technologies used for communication are certain and each of them have their vulnerability against a number of attacks. In \cite{uav_challenges} there is a reference to such attacks against an unmanned vehicle, in general. Thus, even it is a UTM and  is based on   a centralized or decentralized architecture, an eavesdropper has plenty of tools in his hands to attack a UAV solely or its communication channel, with the ability to affect more UTM's nodes. More specific, in case of a decentralized system, the  blockchain's technology complexity implementation allows a  great number of attacks to be applied\cite{blockchain_attacks}. Indicative, some of them are:

\begin{enumerate}
    \item \textit{Liveness Attack:} This attack aims to delay the transaction confirmation time. The attacker tries to gain a potential advantage against honest players to build their private chain. Next is the transaction denial phase in which the attacker attempts to delay the genuine block that contains the transaction and when the attacker decides the delay is unconvincing, then attempt to decrease the rate at which the chain transaction grows.
    \item \textit{Double Spending Attacks:} This kind of harm is generated when one successful transaction is duplicated with the same funds. It represents a potential flaw in digital cash, as the same digital token can be spent two times when such an attack occurs. The conditions allow modified blocks to enter the blockchain. If this happens, the person that initiated the alteration can reclaim sources.
    \item \textit{$51\%$ Vulnerability Attack:} In this  attack is an attack on a cryptocurrency blockchain by a group of miners who control more than $50\%$ of the network's mining hash rate. Owning $51\%$ of the nodes on the network theoretically gives the controlling parties the power to alter the blockchain. 
    \item \textit{Sybil Attack:} An entity on a peer-to-peer network is a piece of software that has access to local resources. An entity advertises itself on the peer-to-peer network by presenting an identity. More than one identity can correspond to a single entity. In other words, the mapping of identities to entities is many to one. Entities in peer-to-peer networks use multiple identities for purposes of redundancy, resource sharing, reliability and integrity. In peer-to-peer networks, the identity is used as an abstraction so that a remote entity can be aware of identities without necessarily knowing the correspondence of identities to local entities.\par
    The Sybil attack in computer security is an attack wherein a reputation system is subverted by creating multiple identities. A reputation system's vulnerability to a Sybil attack depends on how cheaply identities can be generated, the degree to which the reputation system accepts inputs from entities that do not have a chain of trust linking them to a trusted entity, and whether the reputation system treats all entities identically. \par
    In 2018, a successful Sybil attack on Google’s autonomous car led the car to show an incorrect GPS location and caused the vehicle to stop in the middle of the road \cite{sybil_attack}. In this scenario of attack, several fake nodes were successfully added to the network and sent misleading location and traffic condition information to the Google car by exploiting the routing table’s flaws and non-encrypted messages of Google cars.
    
\end{enumerate}

For the above indicative attacks against a blockchain, there is a healthcare scheme that aims to resist in each occasion proposed in \cite{blockchain_attacks}. 

\subsubsection{Side Channel Attacks*}
As mentioned before, besides software attacks, a UAV or even the whole UTM may suffer from hardware attack, such as \textit{Side Channel Attacks}. In general, this kind of  attacks\cite{side_chann_attack_basics}  are a class of physical attacks in which an eavesdropper tries to exploit physical information leakages such as timing information, power consumption, or electromagnetic radiation. Since they are  passive and they can generally be performed using relatively cheap equipment, they are a significant  threat to the security of most cryptographic hardware devices. Such devices may be a personal computers, a small embedded device, smart cards and Radio Frequency Identification Devices (RFIDs). Their introduction in a continuously growing spectrum of applications has turned the physical security and side channel issue into a real concern.\par
Side channel attacks are closely related to the existence of physically observable phenomenons caused by the execution of computing tasks in present microelectronic devices. For example, microprocessors consume time and power to perform their assigned tasks. They also radiate an electromagnetic field, dissipate heat, and even make some electromagnetic noise. A significant part of  digital circuits is based on complementary metal-oxide semiconductors (CMOS). These components used for analog circuits such as image sensors (CMOS sensors), data converters, RF circuits (RF CMOS), and highly integrated transceivers for many types of communication. important characteristics of CMOS devices are high noise immunity and low static power consumption. Since one transistor of the metal–oxide–semiconductor field-effect transistor (MOSFET) pair is always off, the series combination draws significant power only momentarily during switching between on and off states. Consequently, CMOS devices do not produce as much waste heat as other forms of logic, like NMOS logic or transistor–transistor logic (TTL), which normally have some standing current even when not changing state. These characteristics allow CMOS to integrate a high density of logic functions on a chip. It was primarily for this reason that CMOS became the most widely used technology .\par
\begin{figure}[h]
    \centering
    \includegraphics[width=0.7\textwidth]{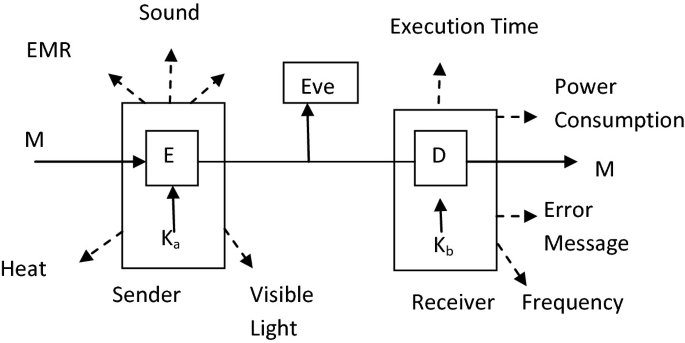}
    \caption{Exploiting power consumption during encryption}
    \label{fig:EMW_in_SCA}
\end{figure}

In side-channel attack though, the eavesdropper is able to monitor the power consumed  during performance of decryption and signature generation. Also, it is possible for eavesdropper to measure the time during performance of cryptographic operation and it is able to analyze how a cryptographic device behaves when certain errors are encountered (fig. \ref{fig:EMW_in_SCA})

In the survey \cite{side_chann_attack} has be done an effort to classify the IoT security attacks. One of them includes the side-channel attack which is sub-classified to several types:
\begin{enumerate}
    \item \textit{Simple Attacks}: The attacker can directly guess the secret key using side-channel information. A simple analysis can help attacker to exploit the relationship between executed operations and the side-channel information
    \item \textit{Differential Attacks}:This attack exploits relationship between side-channel information and processed data.
    \item \textit{Power Analysis Attacks:}
    \begin{enumerate}
        \item \textit{Simple Power Analysis(SPA)}: SPA is a method that calculates directly the power consumption which is collected during encryption and decryption operation. It is based on looking at the visual representation. SPA are even  able to reveal  information about the key used for encryption. Furthermore, SPA can be used to retrieve information about the  cryptographic implementations, i.e how many rounds are used during encryption/decryption. It is the simplest form of power analysis. 
        \item \textit{Differential Power Analysis (DPA)}: This  is an attack method which is much more powerful attack than SPA. In addition to large-scale power variations found with SPA, DPA searches for correlations between different traces. There are several different DPAs.
        \item \textit{Correlation Power Analysis (CPA)}: The CPA is a form of DPA. It differs a bit from the difference of means attack when searching for correlations. CPA uses a power model. This model is used to say something about the power consumption given a specific plaintext and key combination. CPA attacks have many models for expressing this. The two most common power models are the Hamming weight and the Hamming distance models
    \end{enumerate}
\end{enumerate}

\subsection{Data Security Against Quantum Computers}\label{ch:Data_sec_quantum}
As mentioned in \ref{sec:dec-arch}, the blockchain architecture bases its function in strong cryptographic schemes when hashing the new data. Considering the new era of quantum computers, the flawless operation of such a system is jeopardised. Thus, it is highly recommended that the decentralized UTM architectures must encapsulate quantum-resistant algorithms in order to secure not only their data but also human lives.\par
In the popular RSA system, the public key is the product $n=pq$ of two secret prime numbers $p$ and $q$. The security of RSA relies critically on the difficulty of finding the factors $p$, $q$ of $n$. However, in 1994, Shor introduced a fast quantum algorithm~\cite{shor} to find the prime factorization of any positive integer $n$.
In general, suppose the order $r$ of some integer $x$: $x<n$. Suppose further that $x$ is even. This is necessary in order that $x^{ r/2}$ be an integer. By definition then, $r$ is the smallest factor of   $x$.\par
Shor's algorithm, which is designed to take advantage of
the inherent potential of quantum, in contrast to classical computers, exploits a factorization method that differs from the trivial, for large key number $q$.
Then, compute the great common divisor between $x^{ r/2}-1$ and $n$, denoted as gcd$(x^{ r/2}-1, n)$. The Euclidean algorithm takes polynomial time to compute the gcd between two numbers. Since the multiplication of the two numbers gcd$(x^{ r/2}-1)$ gcd$(x^{ r/2}+1)=x^r-1 \equiv 0 \:(mod\: n)$, then gcd$(x^{ r/2}-1)$ and gcd$(x^{ r/2}+1)$ will be two factors of $n$. This procedure  fails only if $r$ is odd, in which case $r/2$ is not in
integer or if $x^{ r/2} \equiv  -1 \: (mod\: n)$
The probability that a randomly selected $x < n=pq$ and coprime to $n$ will have an even order $r$, satisfying the aforementioned equations is at least $1-\frac{1}{2^{k-1}}$, where $k$ is the is the number of  distinct odd prime factors of $n$, and at most is $\frac{1}{2}$. Calculating the gcd of a pair of large numbers using classical computers is a straightforward procedure requiring negligible computing time. Therefore, the feasibility of factoring a large $x< n=pq$ via the procedure described  depends primarily on the feasibility of determining the order $r$ of $x$ (mod $n$), for arbitrarily selected $x$. With classical computers this determination requires solving the discrete log problem.\par
Assume that large quantum computers are built and that they function as smoothly
as one could possibly hope. Shor's algorithm and its generalizations will then
completely break asymmetric key algorithms like RSA, DSA, ECDSA and many other popular cryptographic systems. For example, a quantum computer will find an RSA user's secret key at
essentially the same speed that the user can apply the key.\par
That algorithm is not the only application of quantum computers. A quantum searching algorithm,
introduced by Grover in~\cite{grover1996fast}  finds  a 256-bit AES key in only about $2^{128}$ quantum operations given a few known plaintexts encrypted under that key, which means from $N$ operations to $\sqrt{N}$. Users who want to push the attacker's cost significantly higher than $2^{128}$, the original motivation for 256-bit AES will need a cipher with significantly more than a 256-bit key.

\subsubsection{Quantum Resistance Algorithms}

While the RSA cryptographic system, which is considered reliable and robust, the upcoming era of quantum computers provokes skepticism and anxiety, considering the security of data, including digital signatures. Hence, National Institute of Standards and Technology (NIST) have initiated a process to develop and standardize one or more additional public-key cryptographic algorithms, initializing the era of the Post quantum cryptography (PQC). As a first step in this process, NIST requested public comment on draft minimum acceptability requirements, submission requirements, and evaluation criteria for candidate algorithms. Comments received are posted at its website, along with a summary of the changes made as a result of these comments. The purpose of this notice is to announce that nominations for post-quantum candidate algorithms may now be submitted, up until the final deadline of November 30, 2017. The main cause of the submissions was NIST to choose some algorithms for standardization in both signatures and public key cryptography.\par
During the third round the finalists were the first seven and the  the  other eight algorithms was named as "alternatives". The finalists could continue to be reviewed for a consideration to become a standard, at the conclusion of the third round.   Several of these alternate candidates had worse performance than the finalists but might be selected for standardization based on a high confidence in their security. Other candidates had acceptable performance but require additional analysis or other work to inspire sufficient confidence in their security or security rationale. In addition, some alternates were selected based on NIST’s desire for a broader range of hardness assumptions in future post-quantum security standards, their suitability for targeted use cases, or their potential for further improvement.\par
NIST has completed the third round of the Post-Quantum Cryptography (PQC) standardization process which selects public-key cryptographic algorithms to protect information through the advent of quantum computers. A total of four candidate algorithms have been selected for standardization, and four additional algorithms will continue into the fourth round. So, it  recommends two primary algorithms to be implemented for usage: CRYSTALS-KYBER for key-establishment, while CRYSTALS-Dilithium  and Falcon, \cite{crystals+falcon}, for \textit{digital signatures}. In addition, the signature schemes FALCON and SPHINCS+ will also be standardized.\par

In a decentralized scheme, the computations for decision making accomplished in a remote note which is a cloud server. Thus, the UAVs have the potential to reduce their weight. This case allow to consider the implementation  cryptographic system as an embedded   hardware. This implementation may increase the weight of a UAV but guarantee that the corresponding module will always function without flaws, latency and the need of updates.   It is an asymmetric encryption algorithm developed in 1978 by Robert McEliece. It was the first such scheme to use randomization in the encryption process.   The algorithm is based on the hardness of decoding a general linear code (which is known to be NP-hard). For a description of the private key, an error correcting code is selected for which an efficient decoding algorithm is known, and which is able to correct $t$  errors. McEliece with Goppa codes has resisted cryptanalysis so far. The most effective attacks known use information-set decoding algorithms\cite{isd}. \par

The \textit{McEliece} is a cryptographic system \cite{mceliece}, which is one of the finalists at NIST's contest in order to standardise cryptosystems, foreseen the post quantum era. There have been several efforts to hardware this system \cite{hard_mceliece} in order to make it faster and more reliable. Concerning the concept of the on-board functionalities, the more autonomous an aerial vehicle is the more heavy is. In a centralized architecture the main entity has the major management of the UTM. This implies that the UAVs have less on board procedures to execute. Thus, more mass is available to allow miscellaneous functions to be inserted. A hardware version of the McEleiece cryptosystem seems feasible to be installed  on-board. This action enhance the capability of the UAV to defend against cyber attacks. Furthermore, as the distributed concept highlights the capability to keep the safety of the data on a high level, the  centralized architectures which suffer in this sector, can utilize hardware cryptosystems.     

\section{Conclusion}

The telecommunication science has been evolved in an exotic way. Despite the fact that analog communication did pretty well all the previous years,  we witness the trend of digital communication. The key factors are  confidentiality, integrity and many other  are part of the reliability function. Another factor, that distinguish in our modern and fast era, is speed. Safety is a matter of speed. The more time you have to decide your move, the more robust move you do. The more data you have to process, the robust move you do. Consequently, decision-making , relying in AI, need data in short of time in order to extend the learning capability. Artificial intelligence is another entity that must take a lot of consideration. Several ethical issues arise. These issues have to deal with the proprietary editions of AI algorithms. If these are not open source or they have learned based on biased data, then their usage consider ineffective. \par
As we tend to use unmanned vehicles more and more, then we create a jammed airspace. Simultaneously, aviation industry has been enlarged and the agencies must focus in order to merge the utilization of the airspace. The manned and the unmanned vehicles must to authenticate themselves  when they approach airports, despite their architecture.\par
When a UTM has been built based on decentralized architecture, then different kind of communication applied. Even if they assume to be more reliable, because they eliminate the single point failure, they are exposed to different kind of attacks. Considering that the next era of computation capacity may force entities to reconsider safe communication. While the quantum computers are in the initial stage, NIST managed to standardise algorithms that ensure reliable communication and authentication.

\bibliographystyle{plain}
\bibliography{main}

\begin{thebibliography}{10}

\bibitem{uav_growth}
Unmanned aerial {VehiclesMarket} size \& share analysis - industry research report - growth trends.
\newblock \url{https://www.mordorintelligence.com/industry-reports/uav-market}.
\newblock Accessed: 2024-1-3.

\bibitem{1919telephony}
{\em Telephony}.
\newblock Number v. 77. Telephone Publishing Corporation, 1919.

\bibitem{Barfield_2020}
page i–ii.
\newblock Cambridge Law Handbooks. Cambridge University Press, 2020.

\bibitem{ai_bias}
Hammaad Adam, Aparna Balagopalan, Emily Alsentzer, Fotini Christia, and Marzyeh Ghassemi.
\newblock Mitigating the impact of biased artificial intelligence in emergency decision-making.
\newblock {\em Communications Medicine}, 2(1):149, 2022.

\bibitem{Ahmed2022-kq}
Faiyaz Ahmed, J~C Mohanta, Anupam Keshari, and Pankaj~Singh Yadav.
\newblock Recent advances in unmanned aerial vehicles: A review.
\newblock {\em Arab. J. Sci. Eng.}, 47(7):7963--7984, April 2022.

\bibitem{crystals+falcon}
Ridwane Aissaoui, Jean-Christophe Deneuville, Christophe Guerber, and Alain Pirovano.
\newblock Authenticating civil uav communications with post-quantum digital signatures.
\newblock In {\em 2023 IEEE/AIAA 42nd Digital Avionics Systems Conference (DASC)}, pages 1--9, 2023.

\bibitem{utsim}
Amjed Al-Mousa, Belal~H Sababha, Nailah Al-Madi, Amro Barghouthi, and Remah Younisse.
\newblock Utsim: A framework and simulator for uav air traffic integration, control, and communication.
\newblock {\em International Journal of Advanced Robotic Systems}, 16(5):1729881419870937, 2019.

\bibitem{sybil_attack}
Mohamed Baza, Mahmoud Nabil, Mohamed~MEA Mahmoud, Niclas Bewermeier, Kemal Fidan, Waleed Alasmary, and Mohamed Abdallah.
\newblock Detecting sybil attacks using proofs of work and location in vanets.
\newblock {\em IEEE Transactions on Dependable and Secure Computing}, 19(1):39--53, 2020.

\bibitem{a2a_ldacs_mode}
Miguel~A. Bellido-Manganell and Michael Schnell.
\newblock Towards modern air-to-air communications: the ldacs a2a mode.
\newblock In {\em 2019 IEEE/AIAA 38th Digital Avionics Systems Conference (DASC)}, pages 1--10, 2019.

\bibitem{analogtodigital}
Mohamed~Slim {Ben Mahmoud}, Alain Pirovano, and Nicolas Larrieu.
\newblock Aeronautical communication transition from analog to digital data: A network security survey.
\newblock {\em Computer Science Review}, 11-12:1--29, 2014.

\bibitem{ai_attacks}
Gueltoum Bendiab, Amina Hameurlaine, Georgios Germanos, Nicholas Kolokotronis, and Stavros Shiaeles.
\newblock Autonomous vehicles security: Challenges and solutions using blockchain and artificial intelligence.
\newblock {\em IEEE Transactions on Intelligent Transportation Systems}, 24(4):3614--3637, 2023.

\bibitem{uav_challenges}
Gueltoum Bendiab, Amina Hameurlaine, Georgios Germanos, Nicholas Kolokotronis, and Stavros Shiaeles.
\newblock Autonomous vehicles security: Challenges and solutions using blockchain and artificial intelligence.
\newblock {\em IEEE Transactions on Intelligent Transportation Systems}, 24(4):3614--3637, 2023.

\bibitem{single_p_f}
Philippe Bertin, Servane Bonjour, and Jean-Marie Bonnin.
\newblock Distributed or centralized mobility?
\newblock In {\em GLOBECOM 2009 - 2009 IEEE Global Telecommunications Conference}, pages 1--6, 2009.

\bibitem{Do-178c}
Benjamin Brosgol.
\newblock Do-178c: the next avionics safety standard.
\newblock In {\em Proceedings of the 2011 ACM Annual International Conference on Special Interest Group on the Ada Programming Language}, SIGAda '11, page 5–6, New York, NY, USA, 2011. Association for Computing Machinery.

\bibitem{hard_mceliece}
Shaofen Chen, Haiyan Lin, Wenjin Huang, and Yihua Huang.
\newblock Hardware design and implementation of classic mceliece post-quantum cryptosystem based on fpga.
\newblock In {\em 2022 IEEE High Performance Extreme Computing Conference (HPEC)}, pages 1--6, 2022.

\bibitem{uav_names}
Konstantinos Dalamagkidis, Kimon Valavanis, and Les Piegl.
\newblock {\em On Integrating Unmanned Aircraft Systems into the National Airspace System}.
\newblock 01 2009.

\bibitem{milit_drones_ai}
Thibault de~Swarte, Omar Boufous, and Paul Escalle.
\newblock Artificial intelligence, ethics and human values: the cases of military drones and companion robots.
\newblock {\em Artificial Life and Robotics}, 24(3):291--296, Sep 2019.

\bibitem{side_chann_attack}
Mampi Devi and Abhishek Majumder.
\newblock Side-channel attack in internet of things: A survey.
\newblock In Jyotsna~K. Mandal, Somnath Mukhopadhyay, and Alak Roy, editors, {\em Applications of Internet of Things}, pages 213--222, Singapore, 2021. Springer Singapore.

\bibitem{floreano2015science}
Dario Floreano and Robert~J Wood.
\newblock Science, technology and the future of small autonomous drones.
\newblock {\em nature}, 521(7553):460--466, 2015.

\bibitem{grover1996fast}
Lov~K. Grover.
\newblock A fast quantum mechanical algorithm for database search, 1996.

\bibitem{utm}
Asma Hamissi and Amine Dhraief.
\newblock A survey on the unmanned aircraft system traffic management.
\newblock {\em ACM Computing Surveys}, 56, 09 2023.

\bibitem{uav_comm_sec}
Daojing He, Sammy Chan, and Mohsen Guizani.
\newblock Communication security of unmanned aerial vehicles.
\newblock {\em IEEE Wireless Communications}, 24(4):134--139, 2017.

\bibitem{C2}
Nozhan Hosseini, Hosseinali Jamal, Jamal Haque, Thomas Magesacher, and David~W. Matolak.
\newblock Uav command and control, navigation and surveillance: A review of potential 5g and satellite systems.
\newblock In {\em 2019 IEEE Aerospace Conference}, pages 1--10, 2019.

\bibitem{isd}
Ghazal Kachigar and Jean-Pierre Tillich.
\newblock Quantum information set decoding algorithms.
\newblock In Tanja Lange and Tsuyoshi Takagi, editors, {\em Post-Quantum Cryptography}, pages 69--89, Cham, 2017. Springer International Publishing.

\bibitem{C2_cell}
Robert~J. Kerczewski, Rafael~D. Apaza, Alan~N. Downey, John Wang, and Konstantin~J. Matheou.
\newblock Assessing c2 communications for uas traffic management.
\newblock In {\em 2018 Integrated Communications, Navigation, Surveillance Conference (ICNS)}, pages 2D3--1--2D3--10, 2018.

\bibitem{u-space}
Tim McCarthy, Lars Pforte, and Rebekah Burke.
\newblock Fundamental elements of an urban utm.
\newblock {\em Aerospace}, 7(7), 2020.

\bibitem{ldacs}
Daniel Mielke, Nils Mäurer, Thomas Gräupl, and Miguel Bellido-Manganell.
\newblock {\em Getting Civil Aviation Ready for the Post Quantum Age with LDACS}.
\newblock 04 2021.

\bibitem{sat}
R.W. Murawski, S.C. Bretmersky, and V.K. Konangi.
\newblock Evaluation of vdl modes in the en-route domain.
\newblock In {\em The 23rd Digital Avionics Systems Conference (IEEE Cat. No.04CH37576)}, volume~1, pages 1.A.2--1.1, 2004.

\bibitem{LDACSMAIN}
Nils Mäurer and Arne Bilzhause.
\newblock A cybersecurity architecture for the l-band digital aeronautical communications system (ldacs).
\newblock 10 2018.

\bibitem{faa}
Abhishek Phadke, Josh Boyd, F.~Antonio Medrano, and Michael Starek.
\newblock Navigating the skies: examining the faa’s remote identification rule for unmanned aircraft systems.
\newblock {\em Drone Systems and Applications}, 11:1--4, 2023.

\bibitem{6g}
Muhammad~Mustafa Qureshi, Muhammad~Tanveer Riaz, Saba Waseem, Muhammad~Abbas Khan, and Sidra Riaz.
\newblock The advancements in 6g technology based on its applications, research challenges and problems: A review.
\newblock In {\em Proceedings of the 27th International Conference on Evaluation and Assessment in Software Engineering}, EASE '23, page 480–486, New York, NY, USA, 2023. Association for Computing Machinery.

\bibitem{ai_in_UTM}
Richard Ryan, Saba Al-Rubaye, and Graham Braithwaite.
\newblock Utm regulatory concerns with machine learning and artificial intelligence.
\newblock In {\em 2022 IEEE/AIAA 41st Digital Avionics Systems Conference (DASC)}, pages 1--5, 2022.

\bibitem{message_categ}
Lukas~Marcel Schalk.
\newblock Communication links for unmanned aircraft systems in very low level airspace.
\newblock In {\em 2017 Integrated Communications, Navigation and Surveillance Conference (ICNS)}, pages 6B2--1--6B2--11, 2017.

\bibitem{mceliece}
Nicolas Sendrier.
\newblock {\em McEliece Public Key Cryptosystem}, pages 767--768.
\newblock Springer US, Boston, MA, 2011.

\bibitem{iot_and_5g}
Kinza Shafique, Bilal~A. Khawaja, Farah Sabir, Sameer Qazi, and Muhammad Mustaqim.
\newblock Internet of things (iot) for next-generation smart systems: A review of current challenges, future trends and prospects for emerging 5g-iot scenarios.
\newblock {\em IEEE Access}, 8:23022--23040, 2020.

\bibitem{shor}
P.W. Shor.
\newblock Algorithms for quantum computation: discrete logarithms and factoring.
\newblock In {\em Proceedings 35th Annual Symposium on Foundations of Computer Science}, pages 124--134, 1994.

\bibitem{blockchain}
Bela Shrimali and Hiren~B. Patel.
\newblock Blockchain state-of-the-art: architecture, use cases, consensus, challenges and opportunities.
\newblock {\em Journal of King Saud University - Computer and Information Sciences}, 34(9):6793--6807, 2022.

\bibitem{blockchain_attacks}
Saurabh Singh, A.~S. M.~Sanwar Hosen, and Byungun Yoon.
\newblock Blockchain security attacks, challenges, and solutions for the future distributed iot network.
\newblock {\em IEEE Access}, 9:13938--13959, 2021.

\bibitem{side_chann_attack_basics}
Fran{\c{c}}ois-Xavier Standaert.
\newblock {\em Introduction to Side-Channel Attacks}, pages 27--42.
\newblock Springer US, Boston, MA, 2010.

\bibitem{do-254}
Prashant~S. Vadgaonkar and Ullas Janardhan.
\newblock Do-254/ed-80 - an application guidelines to redesign/re-engineering airborne electronic hardware.
\newblock In {\em SAE 2016 Aerospace Systems and Technology Conference}. SAE International, sep 2016.

\end{thebibliography}

\end{document}